\documentclass[acmsmall]{acmart}

\newcommand{\para }[1]{\medskip \noindent  {\bf #1}}
\newcommand{\1}{{\em (i)}}
\newcommand{\2}{{\em (ii)}}
\newcommand{\3}{{\em (iii)}}

\usepackage{xspace}
\usepackage{subcaption}
\usepackage{booktabs}
\usepackage{multirow}
\usepackage{graphicx}
\usepackage{xcolor}
\usepackage{url}

\newtheorem{obs}{Observation}

\newcommand{\ra}[1]{\renewcommand{\arraystretch}{#1}}

\setcopyright{none}
\renewcommand\footnotetextcopyrightpermission[1]{}
\renewcommand
\begin{document}

\title{Characterizing Concurrency Mechanisms for NVIDIA GPUs under Deep Learning Workloads}

\author{Guin Gilman}
\affiliation{%
  \institution{Worcester Polytechnic Institute}
  \city{Worcester}
  \state{MA}
  \country{USA}
}
\email{grgilman@wpi.edu}

\author{Robert J. Walls}
\affiliation{%
  \institution{Worcester Polytechnic Institute}
  \city{Worcester}
  \state{MA}
  \country{USA}
}
\email{rjwalls@wpi.edu}

\begin{abstract}
We investigate the performance of the concurrency mechanisms available on 
NVIDIA's new Ampere GPU microarchitecture under deep learning training and inference workloads. In 
contrast to previous studies that treat the GPU as a black box, we 
examine scheduling at the microarchitectural level. We find 
that the lack of fine-grained preemption mechanisms, robust
task prioritization options, and contention-aware thread block
placement policies limits the effectiveness of NVIDIA's concurrency
mechanisms.
In summary, the sequential nature of deep learning workloads and their fluctuating resource 
requirements and kernel runtimes make executing such workloads while maintaining 
consistently high utilization and low, predictable turnaround times 
difficult on current NVIDIA hardware.
\end{abstract}
 
\maketitle

\section{Introduction}
\label{sec:intro}

Hazelwood et al. observed that at Facebook data centers, variations in user 
activity (e.g. due to diurnal load) resulted in low utilization periods with 
large pools of idle resources~\cite{facebook}. To make use of these resources, 
they proposed using machine learning training tasks. Analagous low-utilization 
periods have also been observed at the scale of individual GPUs when using both
GPU-based inference~\cite{gslice} and training~\cite{antman}. The proposed solution to this latter problem
was colocating additional inference or training tasks on a single GPU. We go a step further
than these previous studies by considering the GPU at the microarchitectural level rather than treating it as a black box. 
Broadly, we consider the following question: are current GPU application- and 
block-level scheduling mechanisms sufficient to guarantee predictable and 
low turnaround times for latency-sensitive inference requests, while 
also consistently making use of unoccupied resources for best-effort training 
tasks? To answer this question, we explore both NVIDIA's concurrency mechanisms and the characteristics of the workload itself.
Complicating our analyses, the NVIDIA scheduling hierarchy is proprietary and 
some mechanisms (e.g., time-slicing) are not well-documented, so their behavior must be 
reverse-engineered from empirical observation.

We focus on three application concurrency
mechanisms currently offered by NVIDIA devices on the new Ampere 
microarchitecture: priority streams, time-slicing, and multi-process service 
(MPS). We find that all three have important limitations.
For example, when using priority streams, the kernels of the higher-priority 
inference task frequently experience compounded delay as
they are forced to wait behind blocks of training task kernels for GPU 
resources. Time-slicing disallows separate applications from being executed on 
the GPU simultaneously, making it difficult to improve utilization
from a serial execution case. MPS makes it possible to assign a proportional 
share of resources to each application, but it is not possible to assign a 
scheduling priority to a task.

With these limitations in mind, we conclude that a fine-grained block-level
preemption mechanism, if implemented, would improve turnaround time and 
utilization for concurrent deep learning workloads. Such a mechanism would allow the
GPU to preempt any particular subset of thread blocks during their execution to
be resumed at a later point in time. This ability to preempt at the thread-block 
level could be used in conjunction with thread block placement 
policies to improve predictability when servicing inference 
requests, e.g., by choosing placements which minimize resource contention. We 
additionally demonstrate that there
are many opportunities to hide the cost of fine-grained preemption.

Our efforts differ from much prior work in that the analysis presented in this
work is specifically tailored to the use case of concurrent deep learning workloads,
where the inference tasks are latency-sensitive and the training tasks are best-effort. 
We observed that such workloads have fluctuating resource requirements, variable kernel 
runtimes, and sequential kernel launches, and unpredictable arrival times.
Previously proposed thread-block-level scheduling policies~\cite{spatial-multitasking,
warped-slicer, jain2018dynamic, laius, enabling-preemption, chimera} focus only on more generic workloads which 
do not possess such characteristics.
Finally, we add to previous understandings of the CUDA scheduling 
hierarchy and its concurrency mechanisms~\cite{scheduler-details, streams, hierarchy, deadline}.
For example, our observations suggest that resources such as shared memory and registers are not
 transferred on and off the SM between time slices, potentially to reduce the overhead of context-switching.

The remainder of this paper is structured as follows.
Section~\ref{sec:background} provides a description of the CUDA programming
model, as well as introductions to the three concurrency techniques examined in
this work. Our measurement methodology and workload characteristics are described in
Section~\ref{sec:method}. We analyze the performance of the three concurrency 
techniques available on NVIDIA devices in
Section~\ref{sec:mechanisms}. 
In Section~\ref{sec:observations}, we detail a number of key observations which 
demonstrate the potential utility of 
features such as fine-grained preemption on NVIDIA GPUs.
We discuss related work in Section~\ref{sec:related}, and we conclude in
Section~\ref{sec:conclusion}.

 \section{Background}
\label{sec:background}

The following section provides a brief overview of the CUDA programming model
for GPU computing on NVIDIA devices of the Ampere~\cite{ampere}
microarchitecture. It also explains our choice of the three techniques
currently available for executing multiple applications concurrently on the NVIDIA Geforce RTX 3090 GPU:
priority streams, time-slicing, and MPS.

\subsection{CUDA Programming Model}

We limit our description of the programming model to only those details 
necessary to understand concurrent application execution and any performance 
implications thereof. 

\textbf{Kernels, Thread Blocks, Grids, and Warps.} A \emph{kernel} in CUDA programming
is the term for the code which is executed on the GPU. For example, an inference task
is actually a sequence of thousands of kernels executing serially; one kernel might 
compute a single fast-fourier transformation. A kernel is comprised of
a logical array (i.e., a \emph{grid}) of independent \emph{thread blocks}, 
that each execute the same block of code in parallel on different 
subsets of data. A \emph{warp} is a group of 32 threads within a block that 
execute in parallel on the GPU, and 
instructions are issued per warp.

\textbf{Streaming Multiprocessors.} To execute a kernel, its thread blocks are
scheduled to the GPU's \emph{streaming multiprocessors}, or SMs, which are its
hardware units of computation. Each SM in a GPU from the Ampere architecture
has four warp-scheduler units, which can each issue instructions to a warp 
every two cycles~\cite{ampere}. SMs additionally have a fixed set of 
computational resources 
such as threads, shared memory, and registers. The total resource requirements 
of all blocks scheduled to an SM during execution cannot exceed the hardware 
limit for any one of these available resources. An SM is considered to be 
\emph{saturated} if it can schedule no further blocks due to a lack of the
required resources. We consider two blocks to be \emph{colocated} if they 
are executing concurrently on the same SM. 

\textbf{Memory Hierarchy.} Discrete GPUs use a memory hierarchy consisting of
registers, shared memory, L1/L2 cache, and global memory. Discrete here means 
that the GPU is a separate device, often connected to the CPU via PCIe. 
Global memory is GDDR6X DRAM. It is roughly equivalent to CPU main memory,
but it is physically located on the GPU. The L2 cache is shared among the SMs,
while the L1 cache, registers, and shared memory are SM-specific resources.

\textbf{Streams.} A \emph{stream} is a sequence of commands that is executed in
the order they were issued. These includes all operations performed on the GPU,
such as data transfers and kernel launches. Multiple streams can exist 
simultaneously within one \emph{CUDA context}. For our purposes, a CUDA context can be thought of as analogous to a CPU 
process and it contains all resources and actions performed within the CUDA 
driver API. All operations from separate streams are asynchronous and independent from
each other, and streams only interact with each other from within the same 
context. When a \emph{kernel dispatch} command is issued to a stream, it 
launches that kernel to be transferred, scheduled, and executed on the GPU.

\textbf{NVIDIA Scheduling Hierarchy.} When more than one application is being
concurrently executed on a single GPU, there are multiple levels of scheduling 
decisions that occur to determine the final execution schedule. 
\emph{Application-level} scheduling includes the order in which work (such as 
kernel execution or memory transfers) from each application will be computed on
the GPU, while the \emph{thread block scheduler} determines the placement of 
thread blocks onto SMs. For each SM, the \emph{warp scheduler} executes thread 
blocks in groups of 32 threads.

\textbf{Thread Block Scheduler.} Once a kernel arrives at the GPU, its thread
blocks are assigned to SMs by the \emph{hardware thread block scheduler.} A new
block is assigned to an SM as soon as it has enough resources available to 
satisfy that block's resource requirements; which block it chooses to schedule
next is determined by the \emph{leftover policy}~\cite{streams, scheduler-details}, 
while the SM it chooses to place the next block on is chosen by the 
\emph{most-room policy}~\cite{perf}.

\subsection{NVIDIA Concurrency Mechanisms}

We refer to executing two independent applications simultaneously on one GPU as
\emph{concurrent application execution}.
We examine three concurrency mechanisms that NVIDIA offers for supporting
concurrent applications: \emph{priority 
streams}, \emph{time-slicing}, and \emph{multi-process server (MPS)}. We
describe each mechanism in detail and characterize them for deep learning 
workloads in Section~\ref{sec:mechanisms}. 

We make a distinction between the term application and the OS notion of a
process. 
Most commonly, each application is contained within its own process.
However, sometimes it may be advantageous 
to place
logically-separate applications into the same process because it allows for
the developer to have limited control over scheduling priorities. This is the
case when using priority streams. In contrast, when using time-slicing or MPS, 
the applications are in separate processes.

When kernels from separate applications are executed at the same time on a 
single GPU, this is referred to as \emph{concurrent kernel execution}. 
Concurrent application execution can include concurrent kernel execution but 
does not necessarily. In particular, MPS and priority streams allow for the
possibility of concurrent kernel execution, but time-slicing does not.

NVIDIA also offers a fourth
technique for application concurrency known as 
Multi-Instance GPU, which partitions a single GPU into up to seven unique and
isolated instances for separate applications. 
However, because this feature is
not available on the Geforce RTX 3090 Ampere GPU that we used for this study, 
so we forgo analysis of it here.

 \section{Workload Design and Characterization}
\label{sec:method}

We considered a concurrent workload consisting of a single deep learning 
training task and sequence of inference tasks. These workloads were designed to
resemble the scenario of an inference server responding to user requests while 
training models with spare resources. We measured three performance metrics: \1
average turnaround time of the inference requests, \2 variation in turnaround 
time, and \3 the execution time of the training task as a proxy metric for 
utilization. The characteristics of the deep learning models we examined are 
outlined in Table~\ref{tb:workloads}. 
All tests were performed on the NVIDIA Geforce 
RTX 3090 GPU of the Ampere microarchitecture, which has 82 SMs, and each SM has 
a limit of 1536 threads, 16 thread blocks, 64 KB in registers, 1024 KB of 
shared memory, 24 GB DRAM, and 6144 KB L2 cache.

\subsection{Methodology}

We examined models from two sources, the first of which was the Tensorflow 
models from the MLPerf training and inference benchmark 
suites~\cite{mlperf}.\footnote{Specifically, we used v1.0 git commit 
8b58587c93af2a5ee67722064f2540a2db15d42f for the inference suite and v0.7 git 
commit 96ef5cabfccfe06e34e54b6484dd3f6b39293b31 for the training suite.}
To maintain benchmark integrity, we restricted ourselves from making any 
modifications to the MLPerf benchmark models. However, this created two 
additional challenges. The first was that we were unable to get some models 
to build for our platform, so we do not include those models in our study. 
The second was that we could not test priority streams, as that would require 
non-trivial modifications to the benchmarks in order to run both tasks from 
within the same process. Thus, we supplemented these results with five Pytorch 
example models~\cite{pytorch-ex}. Having both Tensorflow and Pytorch models 
allowed us to characterize two popular deep learning model frameworks and model used for 
a variety of purposes including image recognition, speech recognition, and 
natural language processing (NLP).

For each experiment, we ran
one training task and one inference task concurrently. We configured the 
training task to run for the entire duration of the experiment, and the batch 
sizes we used were the maximum possible before encountering an out-of-memory 
error. We used two request patterns for the inference tasks. First, we used a
pattern where the request arrival times followed a Poisson process (i.e., 
MLPerf's server mode). Second, we used a pattern where one
request immediately followed the previous (i.e., MLPerf's single stream mode).
We used 500 requests for the former and 5000 requests for the latter so that 
the inference task would take a comparable amount of time regardless of what 
request pattern was used. For the supplemental CNN models, we only used the
single-stream distribution.

We ran both inference and training without any other concurrent tasks as 
a baseline for comparison. For the Pytorch models, each model was run as both the 
training and inference task for each experiment, while for the MLPerf Tensorflow 
models, RNNT was the training task for both BERT and ResNet-34. The only 
modifications necessary were for evaluating priority
streams with the Pytorch models, as this required some small changes to the 
models so that the training and inference tasks were launched from the same 
process on different CUDA streams.

\subsection{Workload Characteristics}

\begin{table*}[ht]
\centering
\setlength{\tabcolsep}{2pt}
\sffamily
\ra{0.7}
\begin{tabular}{@{}lrrrrrrrrr@{}}
\toprule
                 	                & \textbf{Task}     & \textbf{Backend} & \textbf{Batch} & \textbf{Total} & \textbf{Long-Running}  & \textbf{Large}  \\
                 	                & & & \textbf{Size}  & \textbf{Kernels} & \textbf{Kernels} & \textbf{Kernels} 	   \\
                 	                & & & \textbf{(items)}  & & \textbf{(\% of}  & \textbf{(\% of} 	  \\
                 	                & & & & & \textbf{runtime)}  & \textbf{kernels)}  \\
\midrule  
\textbf{ResNet-50}~\cite{resnet}        & Image Rec & Pytorch & & & & \\
\qquad Training    	                & & & 128 & 212999  & 56.63 & 43.71   \\
\qquad Inference                        & & & 1   & 1011603 & ---   & 15.85   \\
\textbf{ResNet-152}~\cite{resnet}       & Image Rec & Pytorch & & & & \\
\qquad Training    	                & & & 64  & 2187832 & 6.72  & 41.63   \\
\qquad Inference                        & & & 1   & 2843433 & ---   & 7.75    \\
\textbf{AlexNet}~\cite{alexnet}         & Image Rec & Pytorch & & & & \\
\qquad Training    	                & & & 256 & 29402   & 3.28  & 57.85   \\
\qquad Inference                        & & & 1   & 220303  & ---   & 2.28    \\
\textbf{VGG-19}~\cite{vgg}              & Image Rec & Pytorch & & & & \\
\qquad Training    	                & & & 64  & 370612  & 41.60 & 70.64   \\
\qquad Inference                        & & & 1   & 463274  & ---   & 48.68   \\
\textbf{DenseNet-201}~\cite{densenet}   & Image Rec & Pytorch & & & & \\
\qquad Training    	                & & & 64  & 3336809 & 6.76  & 35.93   \\
\qquad Inference                        & & & 1   & 3625505 & ---   & 21.55   \\
\midrule
\textbf{ResNet-34}~\cite{mlperf}        & Image Rec  & Tensorflow & 1    & 1850691 & ---   & 2.65  \\
\textbf{BERT}~\cite{mlperf}             & NLP        & Tensorflow & 1    & 645000  & ---   & 60.23 \\
\textbf{RNNT}~\cite{mlperf}             & Speech Rec & Tensorflow & 1024 & 9409063 & 10.21 & 0.80  \\
\bottomrule
\end{tabular}
  \caption{\label{tb:workloads}The deep learning models analyzed, along with their relevant attributes
  to concurrent performance. {\it Note that the long-running column shows the
  proportion of execution time spent on executing long-running kernels, while the
  large kernels columns show the proportion of large kernels to total kernels. 
  Long-running inference kernels were 
  omitted because they involved a negligible number of such kernels. The MLPerf 
  models were only run as either an inference (ResNet-34, BERT) or training 
  task (RNNT).}}
\label{tb:workloads}
\end{table*}

Whether we are considering training or inference, a deep learning model 
consists of a sequence of kernels that are launched onto the GPU serially to 
perform computations on subsets of the data. In Table~\ref{tb:workloads}, we 
summarize some of the main properties of the kernels that comprised each 
training and inference task we examined. Note that the individual kernels in 
terms of both execution time and required resources. We labeled a kernel as 
\textit{long-running} if it took longer than 1ms to run when executed on the 
GPU in isolation. Another important characteristic of each kernel is how many 
GPU resources it requires. We define a kernel as \textit{large} if it has a 
grid of blocks that cannot all fit onto the GPU's SMs at the same time. This 
situation occurs when all of SMs reach a resource limit while some of the 
kernel's blocks remain unscheduled. In other words, once one resource on an SM 
is exhausted, no more blocks of that kernel can be scheduled to that SM, even 
if there are other resources remaining. The first resource to run out is known 
as the \emph{limiting resource} for a kernel~\cite{perf}.

Long-running
kernels occupy GPU resources for a significant amount of time, and so 
mechanisms that lack the ability to interrupt thread blocks mid-execution 
must wait for them to finish before reassigning those resources. Large
kernels may inefficiently occupy GPU resources by preventing further thread
blocks from being scheduled and making use of the non-limiting resources. We
provide examples of these issues in the next section.

Overall, Table~\ref{tb:workloads} shows that a significant portion of the 
runtime of these workloads was spent on executing large kernels from 
either the training or inference tasks. For some models, such as VGG-19 and 
AlexNet, it is also the case that the majority of the training task's runtime 
consisted of executing long-running kernels. These observations, along with 
those made in the next section, lead to our discussion of preemption-based 
scheduling in Section~\ref{sec:observations}. 
However, there was also a significant amount of kernels that
were small and/or short-running; some models,
such as ResNet-34 and RNNT, have almost no large kernels at all. Therefore, 
there are a number of opportunities to improve utilization by colocating blocks of the two applications.

Importantly, as a single training (or inference) task consists of a sequence of
kernels and each of those kernels has resource requirements and runtimes, this
means that the resource requirements of the task will fluctuate over the course
of its execution as kernels of different sizes and runtimes are launched.

 \section{Characterizing Application Concurrency Mechanisms}
\label{sec:mechanisms}

In this section, we empirically examine and characterize the performance of
priority streams, time-slicing, and MPS for running concurrent deep learning 
workloads on NVIDIA GPUs, presenting both their strengths and weaknesses. As 
described in the previous section, we ran two tasks concurrently: an inference 
task which was a series of inference requests, and a training task. The ideal 
outcome for concurrency would be low and predictable turnaround times for the 
inference task with high utilization of the GPU. We use a proxy metric for 
utilization, which is the execution time of the best-effort training 
task, and we discuss this choice further in Section~\ref{sec:observations}.
Table~\ref{tb:attr} summarizes the distinguishing characteristics of each 
concurrency mechanisms, although these are discussed in more detail below.

\begin{table*}[tb]
\centering
\setlength{\tabcolsep}{2pt}
\footnotesize
\ra{1.0}
\sffamily
\ra{1.3}
\begin{tabular}{@{}lccccccccc@{}}
\toprule
                       & \textbf{Separate Processes}      & \textbf{Colocation}      & \textbf{Priorities}       \\
\midrule  
\textbf{Priority Streams}       & \multicolumn{1}{p{3.5cm}|}{\textbf{\emph{No.}} All applications must be launched from within the same process to make use of priority streams.}		          
                       & \multicolumn{1}{p{3.5cm}|}{\textbf{\emph{Yes.}} Kernels launched from separate CUDA streams can be scheduled to the same SM. However, this will only occur when the kernel from the highest-priority stream has no blocks left to schedule.}                   
                       & \multicolumn{1}{p{3.5cm}|}{\textbf{\emph{Yes.}} Priority streams have three separate priority levels, and the thread block scheduler will always choose to schedule blocks of the kernel from the highest priority stream first at any given time.}                \\
\midrule
\textbf{Time-Slicing}           & \multicolumn{1}{p{3.5cm}|}{\textbf{\emph{Yes.}} Two applications launched as separate processes to the same NVIDIA GPU will be scheduled using time-slicing by default.}			  
                       & \multicolumn{1}{p{3.5cm}|}{\textbf{\emph{No.}} When utilizing time-slicing, two kernels from separate processes are never executed on the GPU at the same time.}       	     
                       & \multicolumn{1}{p{3.5cm}|}{\textbf{\emph{No.}} Time-slicing provides no methods for prioritizing the execution of one application over another, such as specifying the time slice length or frequency for any application.}                        \\
\midrule
\textbf{MPS}                    & \multicolumn{1}{p{3.5cm}|}{\textbf{\emph{Yes.}} Once an MPS server is set up for the target GPU, the two applications are launched as separate processes.}		  
                       & \multicolumn{1}{p{3.5cm}|}{\textbf{\emph{Yes.}} While applications are launched from separate processes, the MPS server is able to schedule any kernels' blocks to the same SM, and to execute them on the GPU simlutaneously.}
                       & \multicolumn{1}{p{3.5cm}|}{\textbf{\emph{No.}} While it is possible to limit the maximum number of threads utilized by each application, there is no method for prioritizing the execution of one process over another.}                    \\
\bottomrule
\end{tabular}
\caption{A comparison of the main attributes of each concurrency mechanism:
  their ability to be used on kernels from separate processes, the possibility
  of colocating blocks from different tasks, and whether or not prioritization
  of a specific task is possible.}
\label{tb:attr}
\end{table*}

\begin{figure*}[t]
    \centering
    \begin{subfigure}{0.45\columnwidth}
    	\centering
    	\includegraphics[width=\columnwidth]{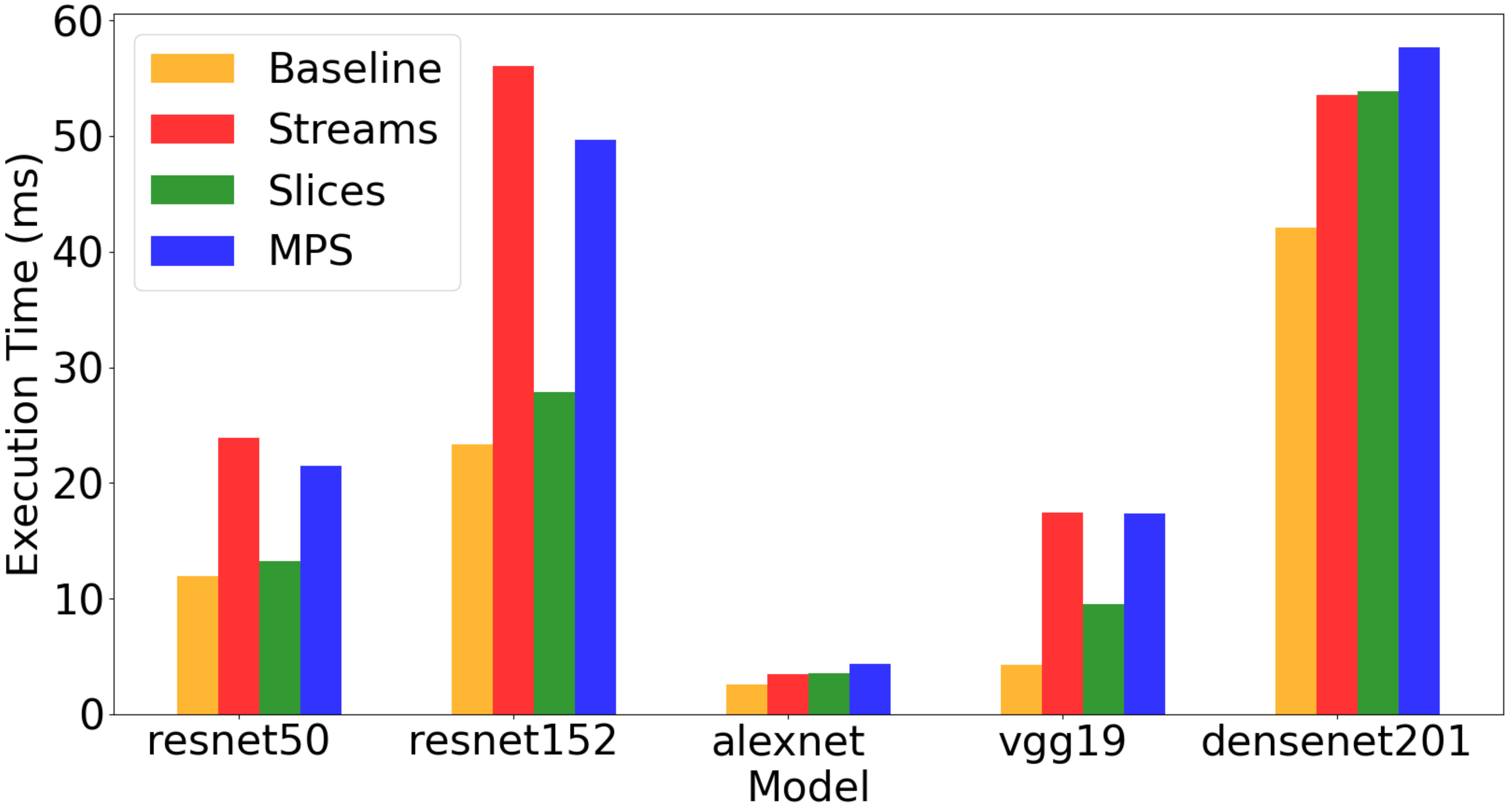}
    	\caption{Average Turnaround Time for Inference}
    	\label{fig:turn}
    \end{subfigure}
    \begin{subfigure}{0.45\columnwidth}
    	\centering
    	\includegraphics[width=\columnwidth]{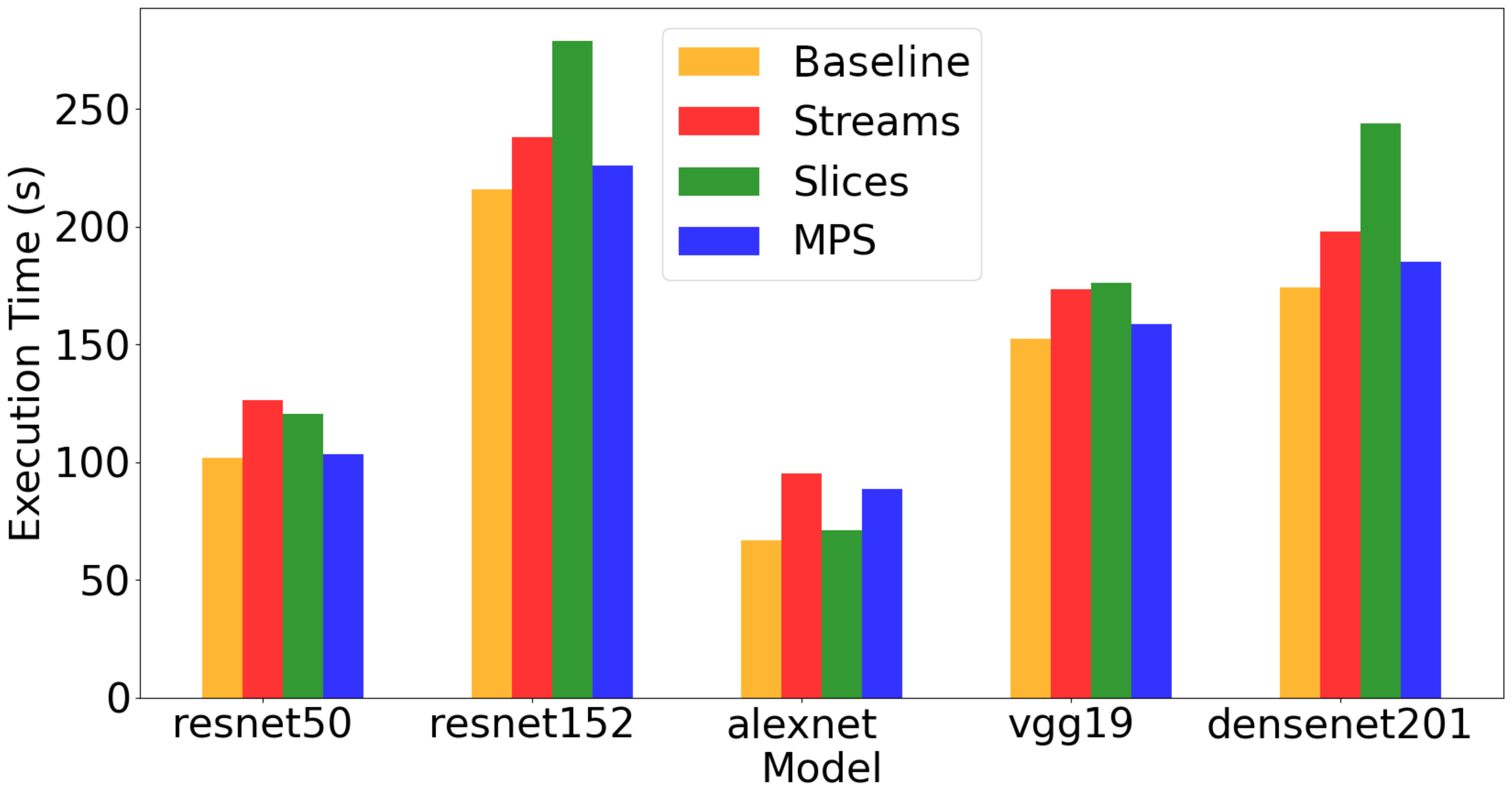}
    	\caption{Average Execution Time for Training}
    	\label{fig:util}
    \end{subfigure}
    \caption{The average turnaround times and utilization for each of the 
    three mechanisms on five different models. {\it Note that the turnaround times
    are the averages of 5000 inference requests in milliseconds, and the measurement of 
    training execution time is the average of 10 runs in seconds. The baseline is the 
    time taken when run in isolation.}}
\label{fig:tt_util}
\end{figure*}

\begin{figure*}[t]
	\centering
    \begin{subfigure}{\columnwidth}
    	\centering
    	\includegraphics[width=\columnwidth]{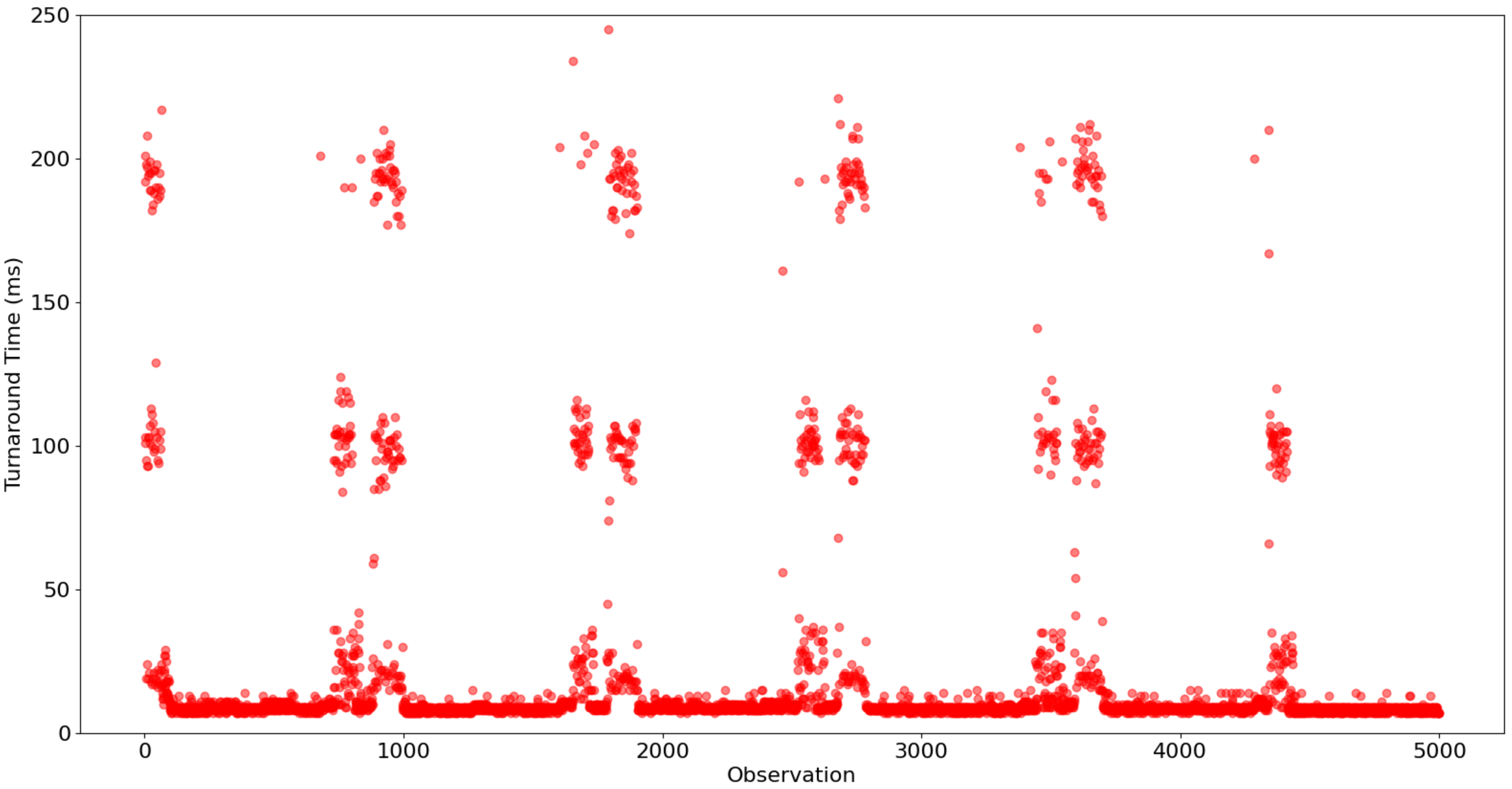}
    	\caption{Priority Streams}
    	\label{fig:var-streams}
    \end{subfigure}
    \begin{subfigure}{0.45\columnwidth}
    	\centering
    	\includegraphics[width=\columnwidth]{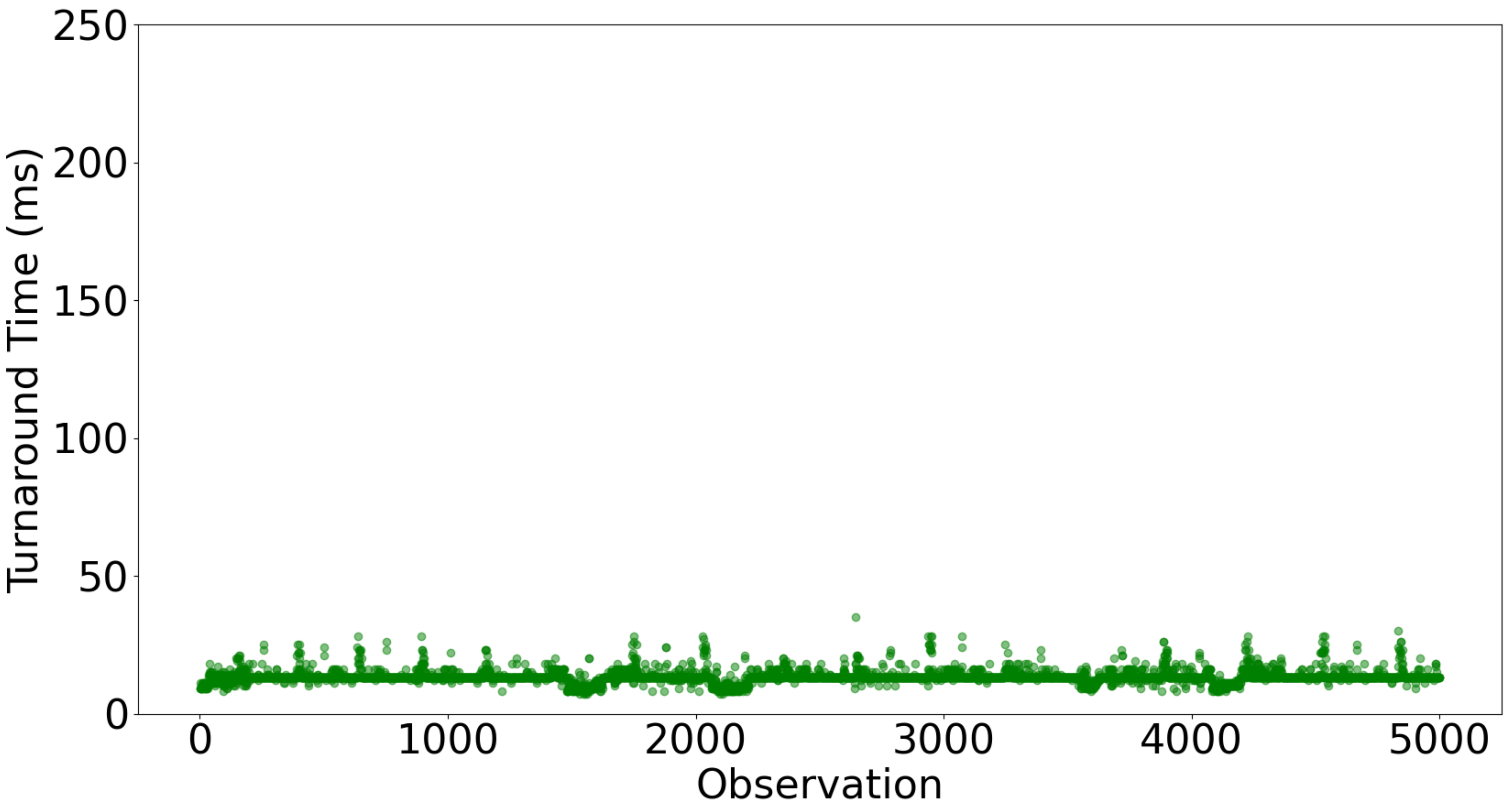}
    	\caption{Time-Slicing}
    	\label{fig:var-slices}
    \end{subfigure}
    \begin{subfigure}{0.45\columnwidth}
    	\centering
    	\includegraphics[width=\columnwidth]{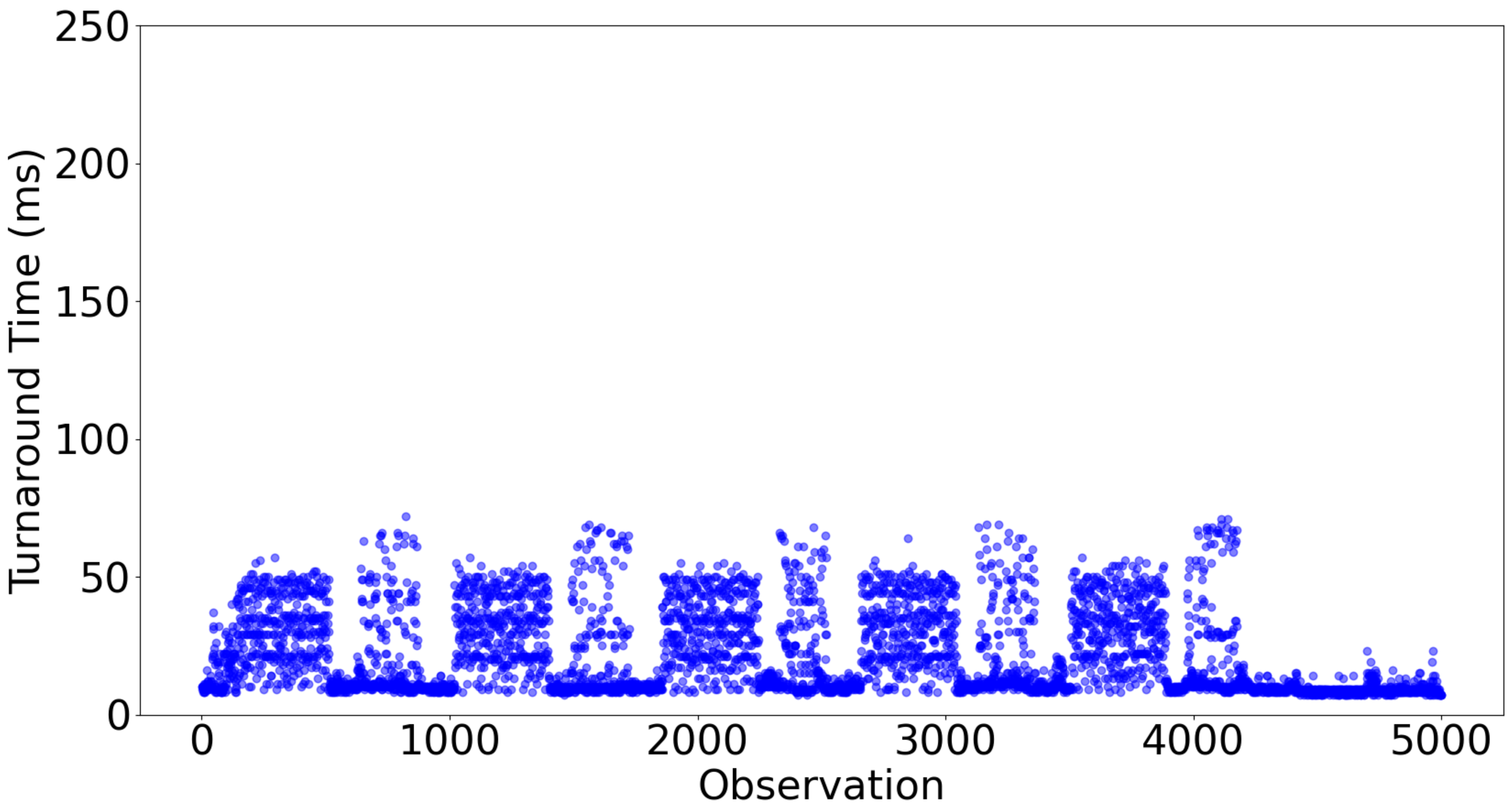}
    	\caption{MPS}
    	\label{fig:var-mps}
    \end{subfigure}
    \caption{The variance of the turnaround times for the ResNet-50 model. 
    {\it Other models' variance results were omitted for space, but resemble 
    these.}}
\label{fig:variance}
\end{figure*}

\begin{figure*}[ht]
    \centering
    \begin{subfigure}{0.45\columnwidth}
    	\centering
    	\includegraphics[width=\columnwidth]{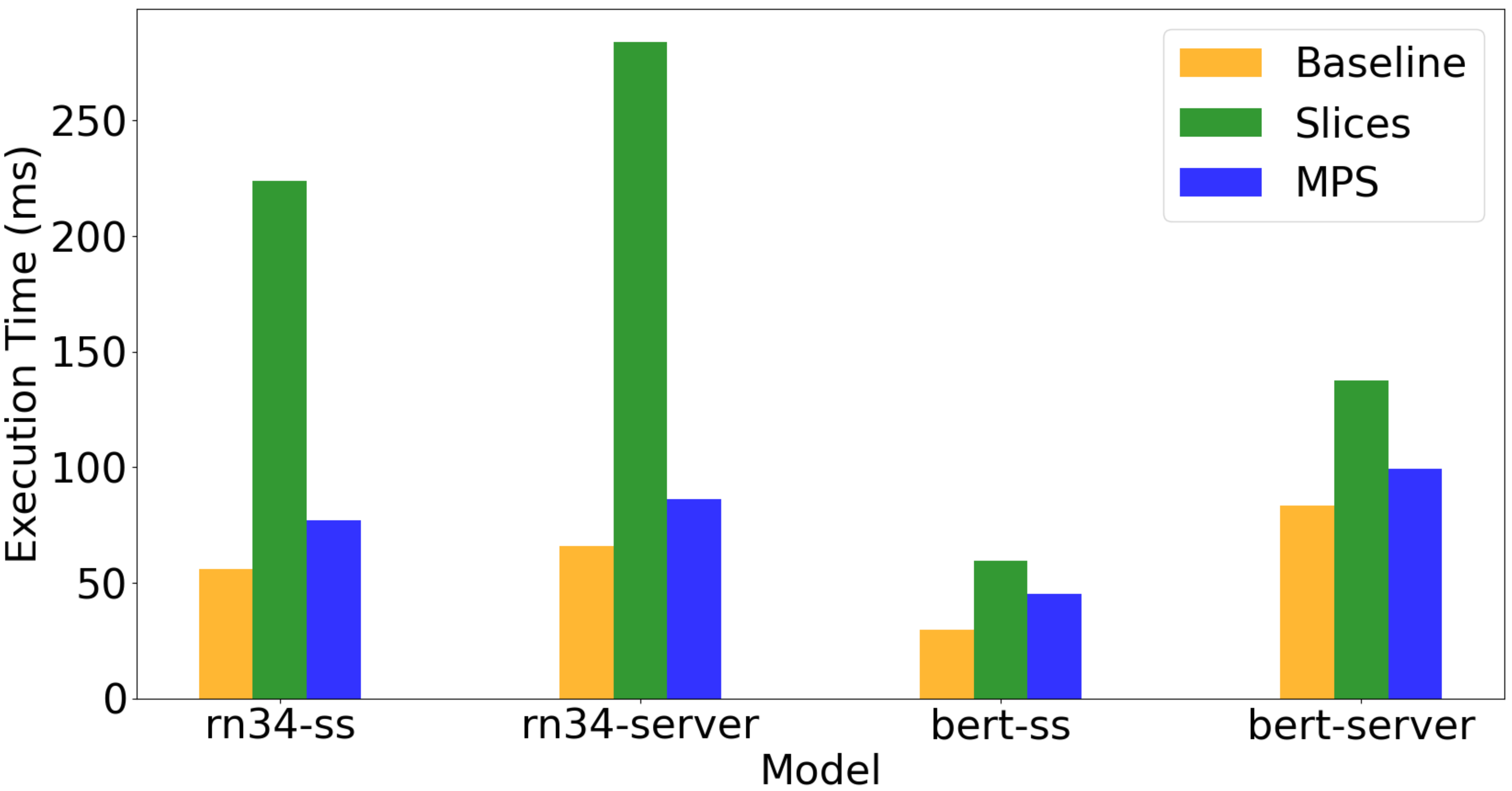}
    	\caption{Average Turnaround Times}
    	\label{fig:turn_rnn}
    \end{subfigure}
    \begin{subfigure}{0.45\columnwidth}
    	\centering
    	\includegraphics[width=\columnwidth]{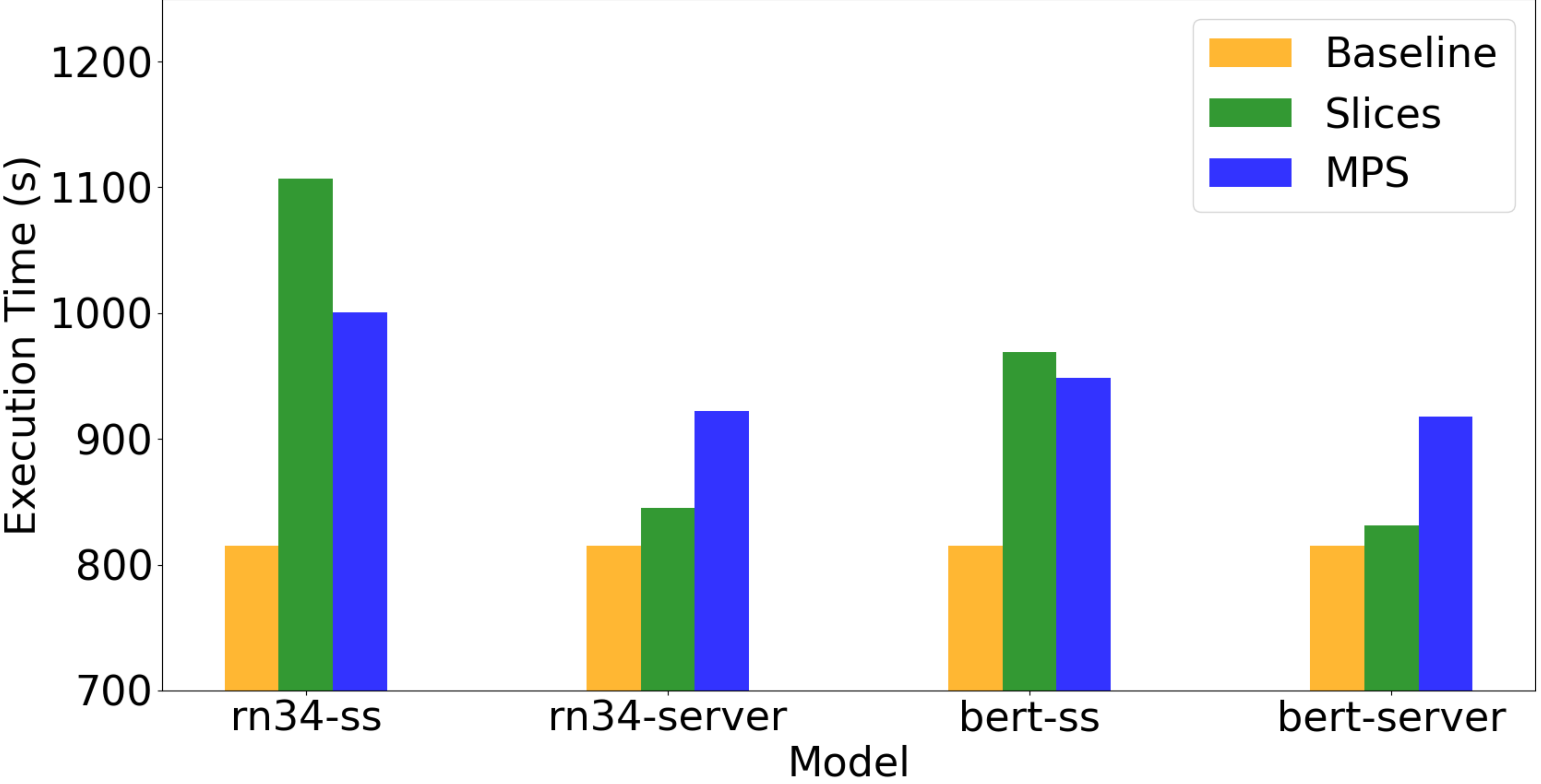}
    	\caption{Average Utilization}
    	\label{fig:util_rnn}
    \end{subfigure}
    \caption{The average turnaround times and utilization for each of the 
    three mechanisms on the MLPerf models. {\it Note that the turnaround times
    are the averages of 5000 consecutive inference requests in milliseconds in the single-stream (ss)
    scenario, and 500 requests which arrive via a Poisson process in the server mode. Additionally,
    the measurement of training execution time is the average of 10 runs in seconds. The baseline is the 
    time taken when run in isolation.}}
\label{fig:rnn_tt_util}
\end{figure*}

\begin{figure*}[ht]
	\centering
    \begin{subfigure}{0.45\columnwidth}
    	\centering
    	\includegraphics[width=\columnwidth]{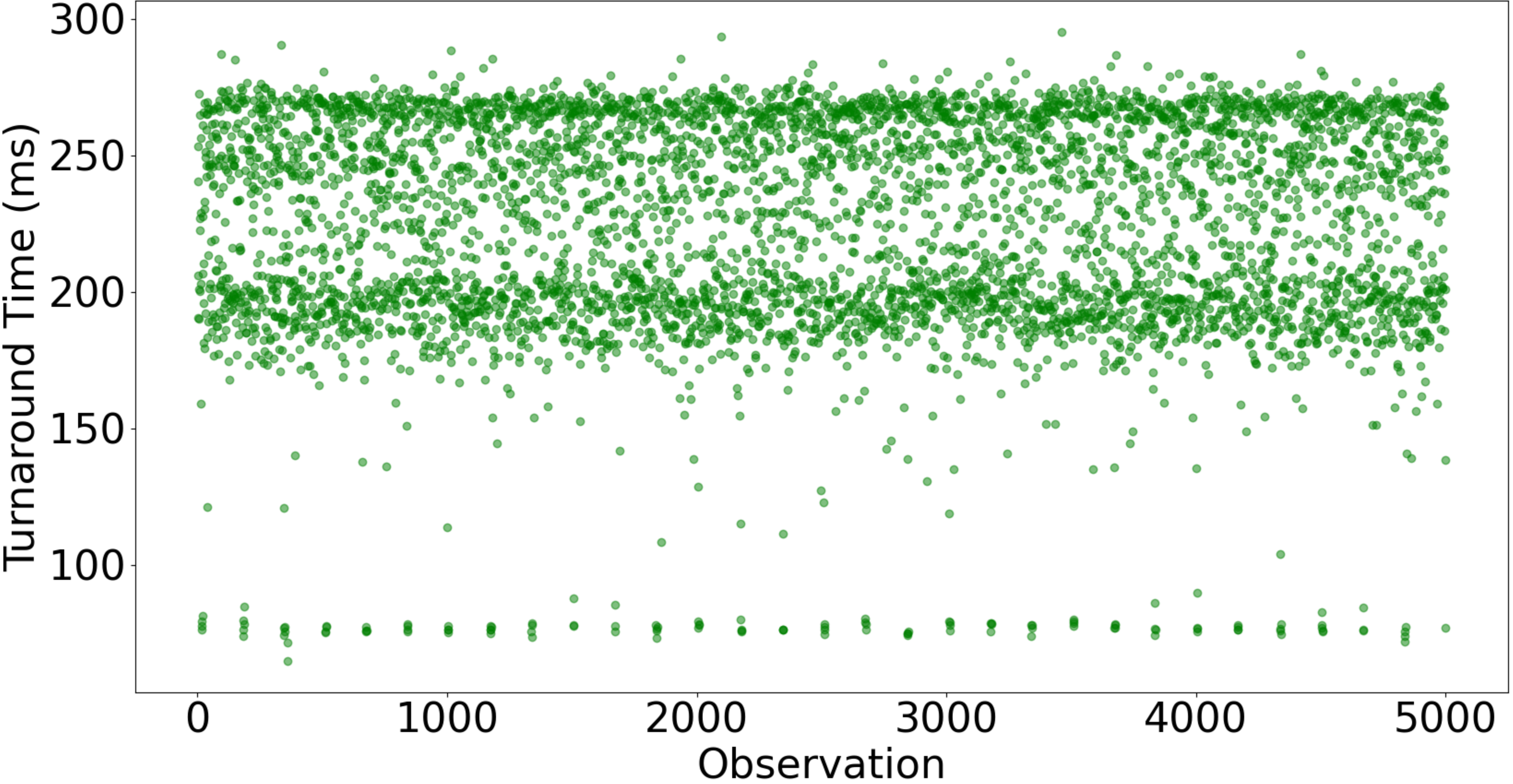}
    	\caption{Time-Slicing}
    	\label{fig:rnn-slices}
    \end{subfigure}
    \begin{subfigure}{0.45\columnwidth}
    	\centering
    	\includegraphics[width=\columnwidth]{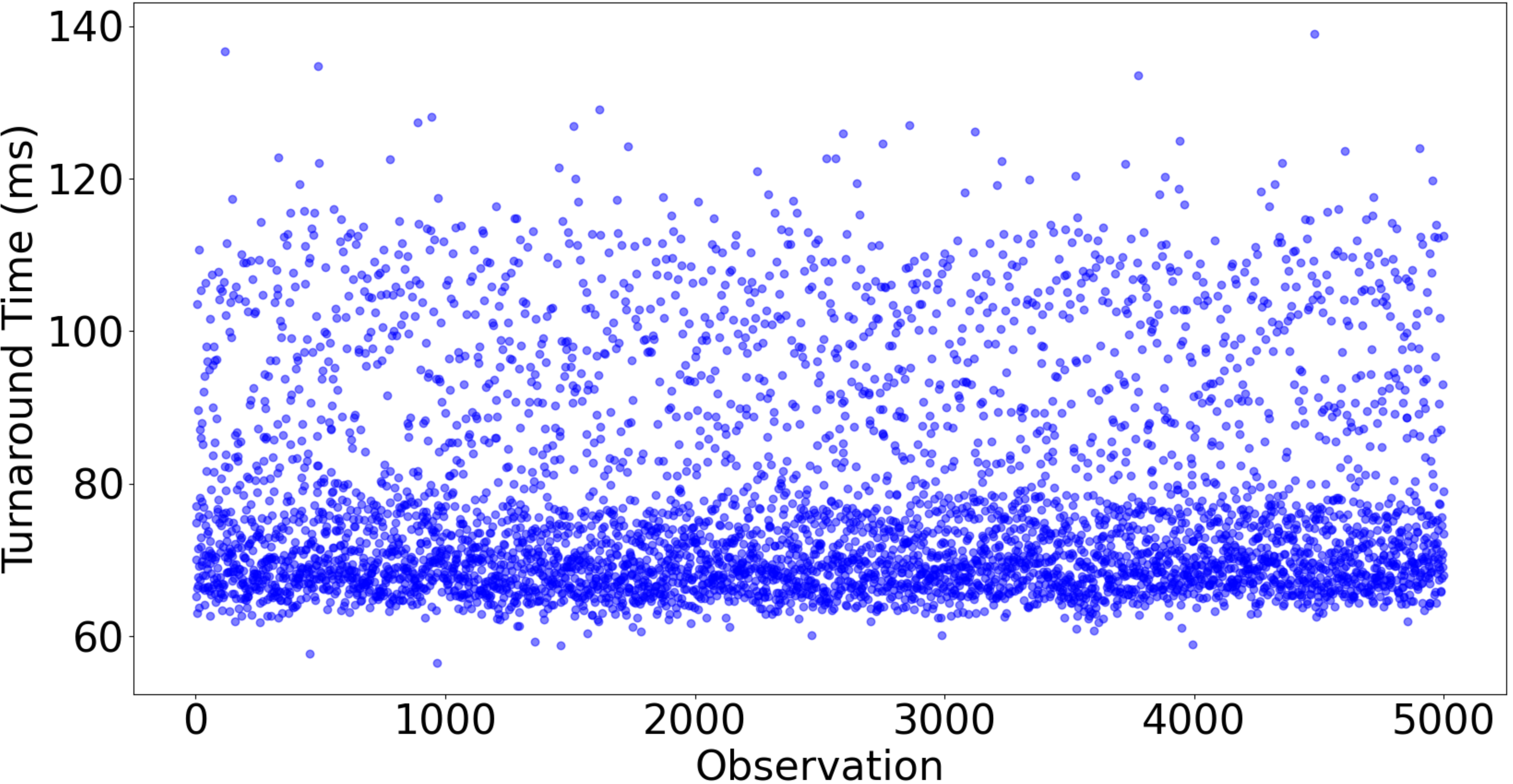}
    	\caption{MPS}
    	\label{fig:rnn-mps}
    \end{subfigure}
    \caption{The variance of the turnaround times for the ResNet-34 model, in
    the consecutive 5000 inference requests scenario. 
    {\it Other models' variance results were omitted for space, but resemble 
    these.}}
\label{fig:ss_variance}
\end{figure*}

\begin{figure*}[t]
	\centering
    \begin{subfigure}{0.45\columnwidth}
    	\centering
    	\includegraphics[width=\columnwidth]{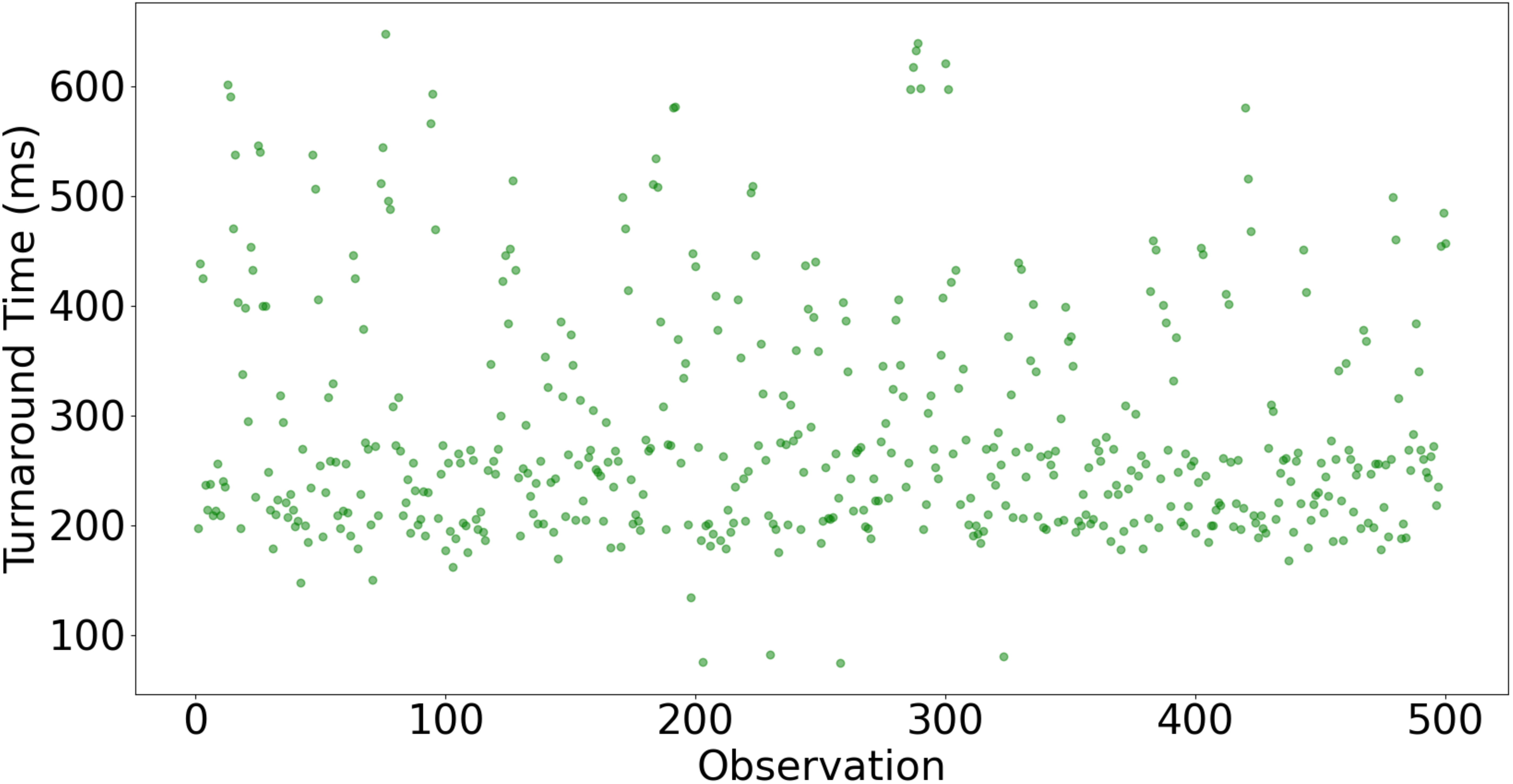}
    	\caption{Time-Slicing}
    	\label{fig:rnn-slices-se}
    \end{subfigure}
    \begin{subfigure}{0.45\columnwidth}
    	\centering
    	\includegraphics[width=\columnwidth]{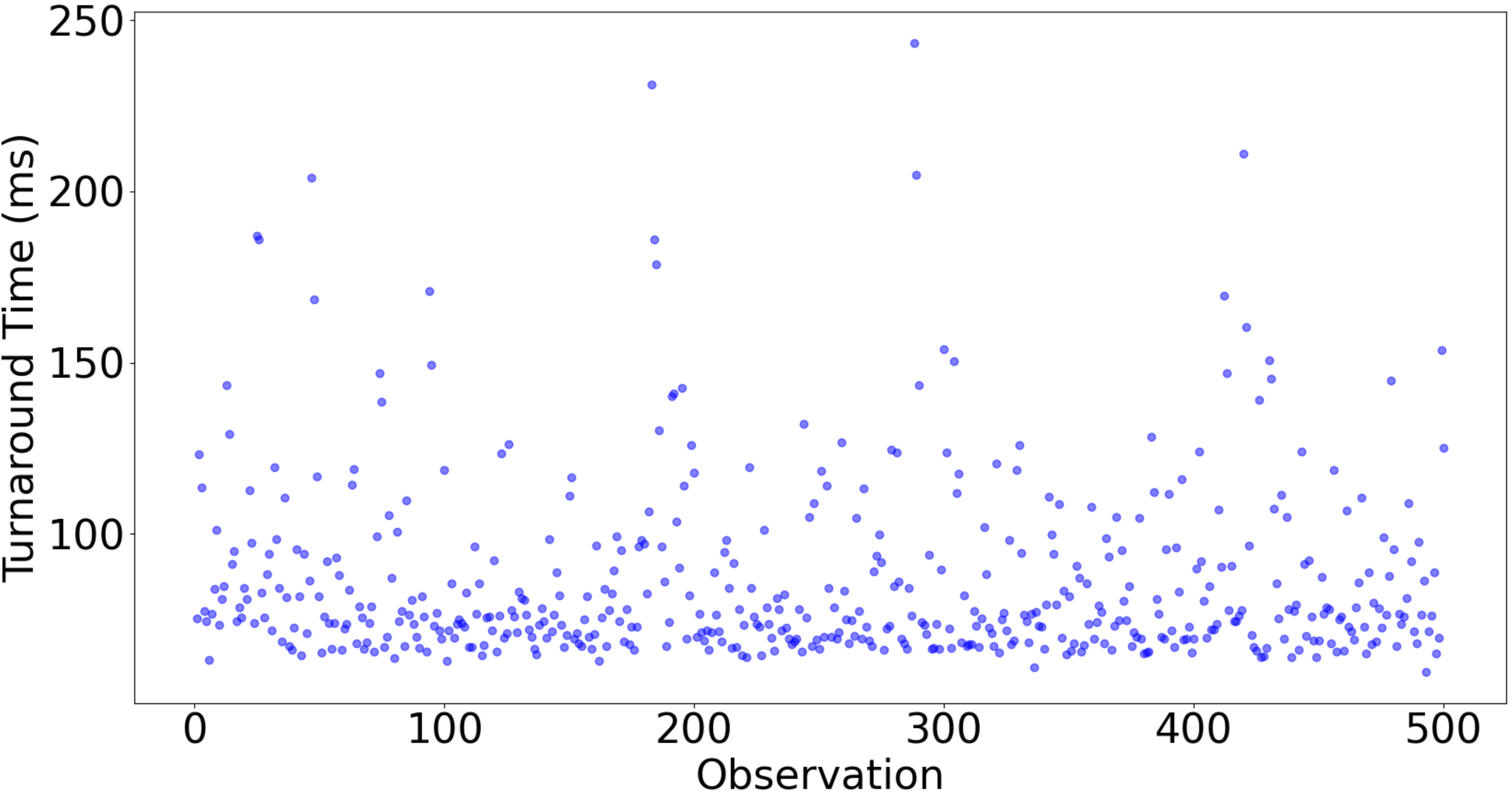}
    	\caption{MPS}
    	\label{fig:rnn-mps-se}
    \end{subfigure}
    \caption{The variance of the turnaround times for the ResNet-34 model, in
    the MLPerf server scenario. 
    {\it Other models' variance results were omitted for space, but resemble 
    these.}}
\label{fig:server_variance}
\end{figure*}

\subsection{Priority Streams}
\label{sec:prio_characterization}

When using priority streams, the kernels of the two applications are launched 
from within the \emph{same process} on different streams using threads. Streams can be 
assigned one of three priorities ranging from -2 to 0. The thread block 
scheduler will always pick blocks from the highest-priority stream first when 
scheduling, but it will not interrupt any thread blocks currently being 
executed on the GPU. We implemented application-level concurrency by putting 
both applications within the same OS process, but launching them onto separate 
CUDA streams, with the kernels of the inference task being on higher-priority 
streams than those of the training task.

\begin{obs}[O\ref{obs:compound}] \label{obs:compound}
Priority streams cannot preempt executing thread blocks in the middle of 
  execution, and this resulted in compounded delay and resource contention 
  leading to high and less predictable turnaround times.
\end{obs}

When a kernel from a higher-priority stream arrives at the GPU,
its thread blocks will take precedence over any \emph{unscheduled} blocks
of any lower-priority kernels. However, the high priority kernel cannot
interrupt the execution of a lower priority kernel's already-executing thread blocks.
In other words, the higher-priority kernel must wait for any currently-executing blocks
from a lower-priority stream to finish.

This led to a phenomenon we term \emph{compounded delay.}\footnote{Compounded 
delay as we have described it here can be though of as an instance of the 
\emph{convoy effect}~\cite{ostep}.} As described in the previous section, all 
of our examined models were structured as a sequence of consecutive kernels. In
our experiments, when a high priority kernel finished executing, there was a 
window of time before the next kernel could reach the GPU. In this timeframe, 
there were no inference kernels ready to execute, so the lower-priority 
training kernel would resume executing and fill the GPU with its thread blocks. 
Shortly after resuming the training kernel, the next inference kernel would 
arrive. As the priority streams mechanism does not support preemption of 
executing thread blocks, the inference kernel had to wait for the 
currently-executing training blocks to finish. 

We can see the effects of this delay in the results from the ResNet-50, 
ResNet-152, VGG-19, and DenseNet-201 models in Figure~\ref{fig:turn}, where 
the turnaround times were approximately 2X, 3X, 4X, and 1.75X compared to the 
baseline, respectively. The impact of the compounded delay was dependent on the
characteristics of the training kernels. For example, the ResNet models and
VGG-19 saw some of the worst turnaround times as the inference task, as these 
models spent about half of their training task's time on executing large or long-running
kernels. Intuitively, 
long-running kernels resulted in more compounded delay. In effect, we can see 
that compounded delay essentially canceled out any benefits one might expect to
gain from placing the inference kernels on a higher-priority stream. In 
particular, priority streams' turnaround times were comparable to that of MPS 
in almost all cases, despite MPS having no notion of priorities.
Compounded delay also reduced the predictability in turnaround time as seen in 
Figure~\ref{fig:var-streams}. We observed spikes in
turnaround time during the time the training epochs were executing on 
the GPU, as the kernels of the inference task are interacting with and being
delayed by those of the training task. 

It is additionally worth noting that some of the performance degradation can be
explained by the effects of resource contention when the thread blocks of the 
two applications are colocated on the same SM. For example, the blocks from 
the two kernels may contend for the SM's warp scheduler. Official documentation
does not describe how the warp scheduler interacts with priority streams. If 
the warp scheduling policy does not prioritize the blocks from the higher 
priority stream, such as if the warp scheduler uses a greedy-then-oldest or
loose round-robin policy~\cite{hierarchy}, then the warp scheduler would 
effectively de-prioritize the higher priority thread blocks.

\subsection{Time-Slicing}

When two applications are run as separate processes and MPS is not being used, 
the CUDA application-level scheduler will alternate between the processes over 
time, yielding the GPU's computational resources (e.g., the warp scheduler and 
computational units) completely to one process for the duration of a 
time-slice~\cite{deadline, hierarchy}. Only one application's kernels are
executing on the GPU during any given time slice. We implemented
application-level concurrency by launching each application as a separate 
process.

\begin{obs}[O\ref{obs:slicing}] \label{obs:slicing}
Time-slicing tended to exhibit predictable and low turnaround times for
  models with relatively low baseline turnaround times (unless there is memory 
  transfer contention; see Observation~\ref{obs:transfers}), due to a lack of 
  interference from the training task. This came at the cost of poor 
  utilization, as the two tasks never actually executed on the GPU at the same 
  time.
\end{obs}

As demonstrated in Figure~\ref{fig:var-slices}, time-slicing offers the most 
predictable performance of the three. We attribute this high predictability to 
two factors. The first is that the training and inference tasks never execute 
at the same time, so there is no contention for SM resources during block 
execution. Second, executing blocks can be preempted (although this preemption 
is coarse-grained and yields the entire GPU to the preempting task), so the 
inference kernel does not need to wait for any blocks of the training task to 
finish executing before being scheduled to the GPU, i.e. compounded delay is 
not a problem. 

The primary factor that influences turnaround time is the number of other GPU 
applications that are being executed concurrently, as this changes the amount 
of time that any one job must wait for access to the GPU's resources. The 
reason for this is that time slices are a fixed size and are assigned 
round-robin to each process. The precise behavior of the time-slicing mechanism
is not well-documented. However, we determined in our empirical setup that time
slices were fixed-length and assigned round-robin across processes. 
Empirically, we determined that the time slice length is fixed to approximately
2ms. These observations are consistent with those made in previous work about 
the earlier Turing microarchitecture~\cite{deadline, hierarchy}. As far as we 
could determine, the time slice size and priority assigned to each process 
cannot be configured.\footnote{Jetson devices do allow time slice 
configuration~\cite{deadline}.} This means that there is no way for one 
application to be prioritized over another, either by extending an 
application's time on the GPU or the frequency with which their time slices 
are scheduled.

Therefore, the trade-off inherent in using time-slicing is predictability at 
the cost of utilization, which was frequently the worst of the three surveyed 
mechanisms. In particular, extra computational GPU resources remain idle during
each time slice. 
For kernels which do not fully occupy the GPU in terms of threads, registers, 
shared memory, and grid size, time-slicing does nothing to occupy those 
resources.

The lack of spatial sharing is the major limitation of time-slicing, as it does
not truly solve the GPU resource utilization problem being addressed by
concurrently executing training and inference tasks. For the ResNet models and 
particularly for DenseNet-201, the training time increased to over 100 seconds 
more than the baseline in Figure~\ref{fig:turn}. The reason VGG-19 and 
AlexNet did not see such an increase is due to the shorter lengths of their 
inference tasks; they completed earlier, allowing the training task to then 
utilize the GPU resources in isolation.

\begin{obs}[O\ref{obs:contextswitch}] \label{obs:contextswitch}
Time-slicing is limited by the fact that the two tasks can only be launched 
  together if the sum total of the resources required by both is less than the 
  total available on the GPU, despite the fact that they never execute on
  the GPU at the same time.
\end{obs}

While the inference task is the only application executing during its time 
slice, our observations suggest that it still has to share certain resources
such as registers and shared memory with the other process. We determined 
empirically that the resource requirements of any tasks being run 
simultaneously as separate processes cannot together exceed the resource 
limitations of the GPU, or an error will be thrown. For instance, the GPU had 
64KB of registers per SM. When we launched two applications that each used 40KB
of registers per block, with exactly enough blocks for one per SM, it caused 
the second process to reach the GPU for scheduling to crash with an
out-of-memory error. We empirically observed similar behavior for shared and
global memory, as well. We hypothesize that this is the case because resources 
such as shared memory, registers, and global memory that are used by a process
are not transferred on and off the GPU between time slices, and we suspect the 
reason is to avoid prohibitively high context-switching overheads when 
swapping between applications across time slices. 

This has performance implications for the training task, which had to be scaled
down from its maximum batch size in order to allow space for the inference 
task. In other words, despite the fact that the two tasks never execute on the 
GPU at the same time, neither task can fully utilize the GPU during their time 
slice without running into this error. 
The implication is that we have to be conservative when setting the 
batch size for the training task, as that is what determines how many resources 
each individual kernel is going to use. Since it cannot be known a priori which
inference kernels will be running alongside which training kernels, it must be 
based on the worst-case scenario.

This problem is compounded by the fact that in some systems, it may be 
difficult to know ahead of time precisely how many resources the inference task
will require. Given that the rate at which inference requests will be received
will be unknown, we can either perform inference for a single image at a time, 
choose a fixed batch size to use for performing inference, or perform 
inference using variable-sized batches. Single image inference has
predictable resource usage, so an out-of-memory error can be avoided; however,
this will add queueing delay (i.e., one image now has to wait for the
previous request to be serviced first). Fixed batch sizes also have predictable
resource usage as they are just the generalized case of single-image inference,
so we can tune the training task to accommodate that while minimizing queueing 
delay. However, if we do not fill up the batch for a particular run, then we 
will have even lower utilization.

\begin{obs}[O\ref{obs:transfers}] \label{obs:transfers}
Contention due to memory transfers can adversely impact predictability and
  turnaround time.
\end{obs}

\begin{figure*}[t]
    \centering
    \begin{subfigure}{0.45\columnwidth}
    	\centering
    	\includegraphics[width=\columnwidth]{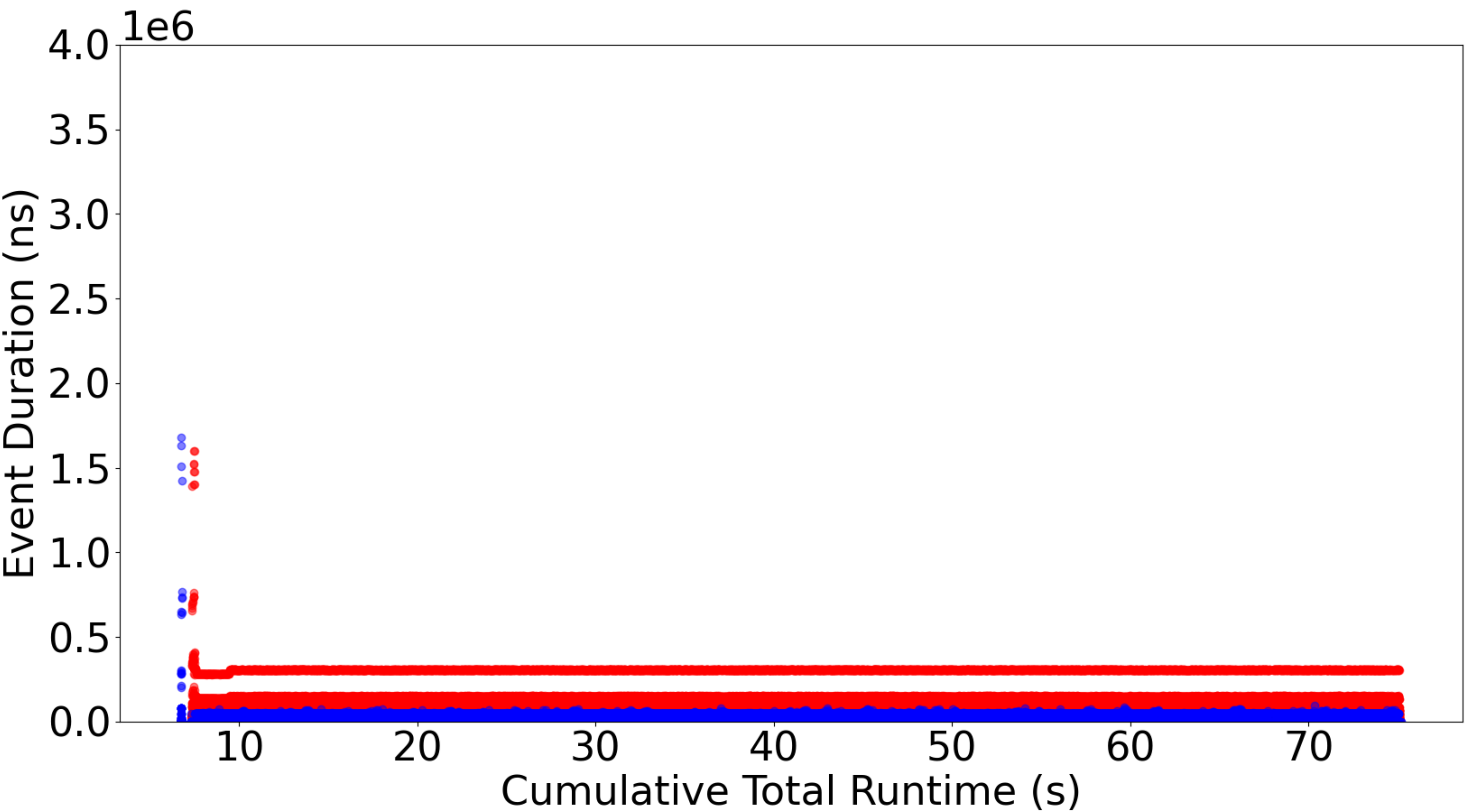}
    	\caption{ResNet-34 Kernel and Transfer Times (Baseline)}
    	\label{fig:transfer_base}
    \end{subfigure}
    \begin{subfigure}{0.45\columnwidth}
    	\centering
    	\includegraphics[width=\columnwidth]{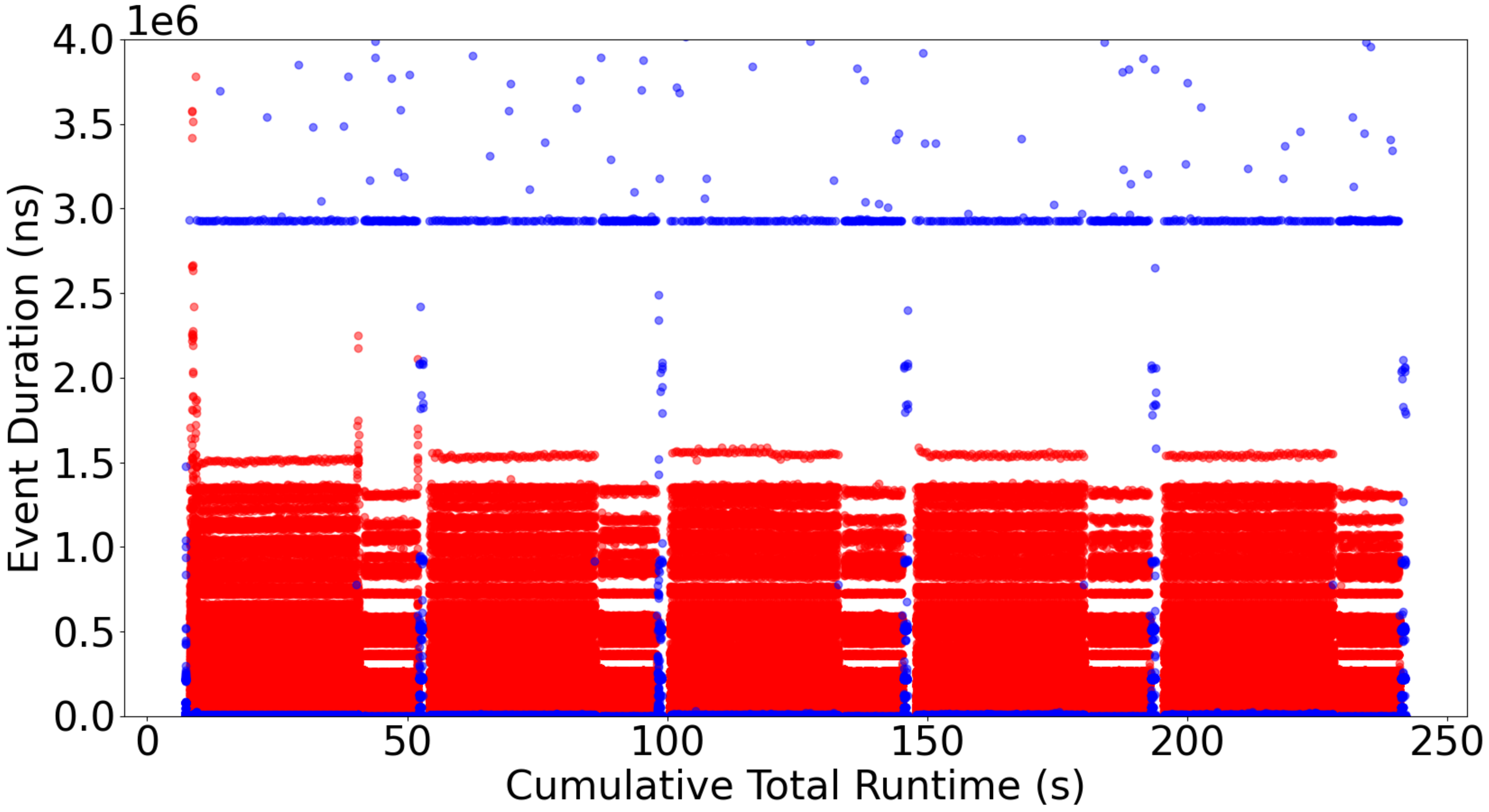}
    	\caption{ResNet-34 Kernel and Transfer Times (Time-Slicing)}
    	\label{fig:transfer_slice}
    \end{subfigure}
    \caption{Kernel execution times (red) and memory transfer operation times 
    (blue) for the ResNet-34 inference task in both the baseline and time-slicing
    scenarios.}
    \label{fig:transfer-res}
\end{figure*}

\begin{figure*}[t]
    \centering
    \begin{subfigure}{0.45\columnwidth}
    	\centering
    	\includegraphics[width=\columnwidth]{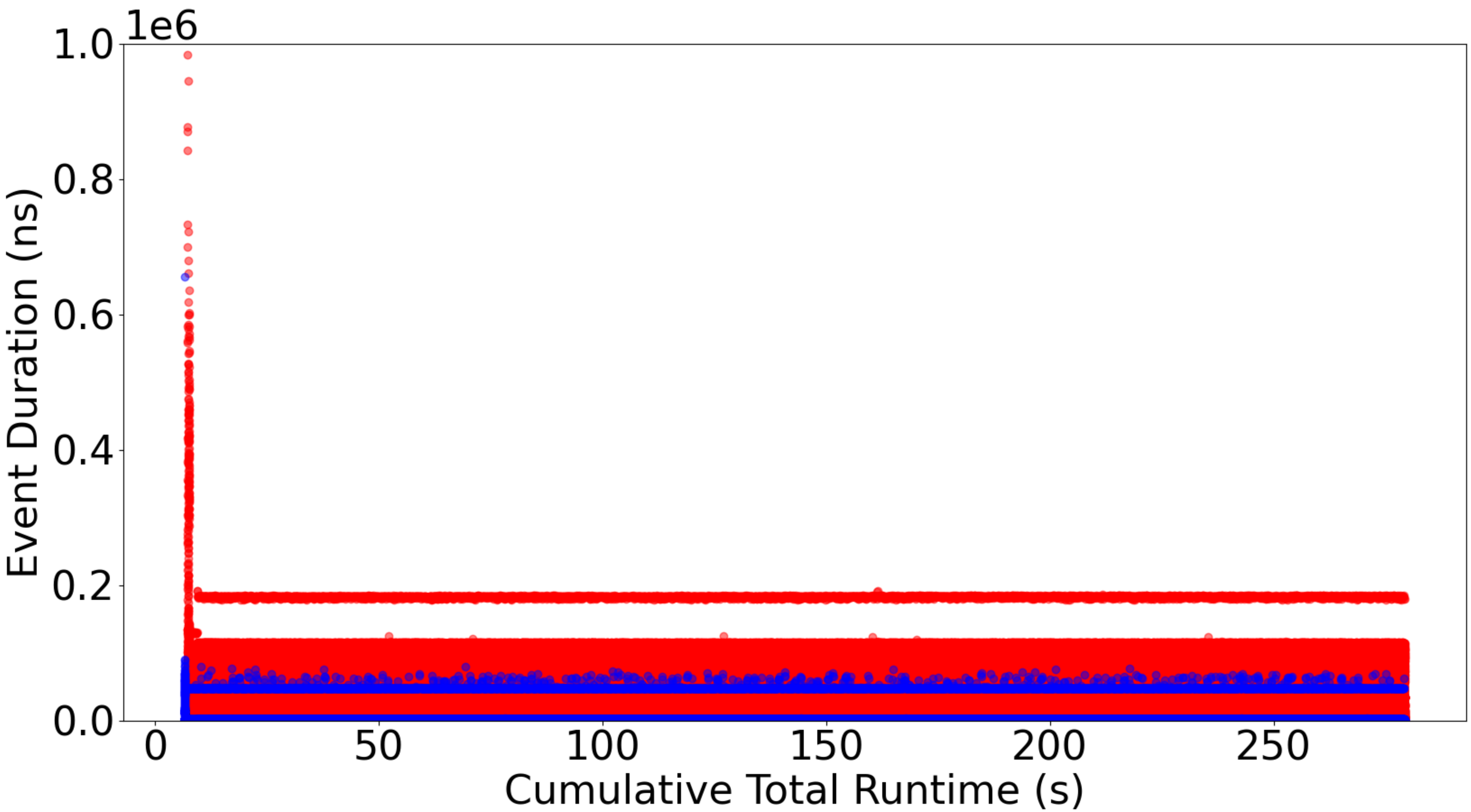}
    	\caption{DenseNet-201 Kernel and Transfer Times (Baseline)}
    	\label{fig:transfer_base_dense}
    \end{subfigure}
    \begin{subfigure}{0.45\columnwidth}
    	\centering
    	\includegraphics[width=\columnwidth]{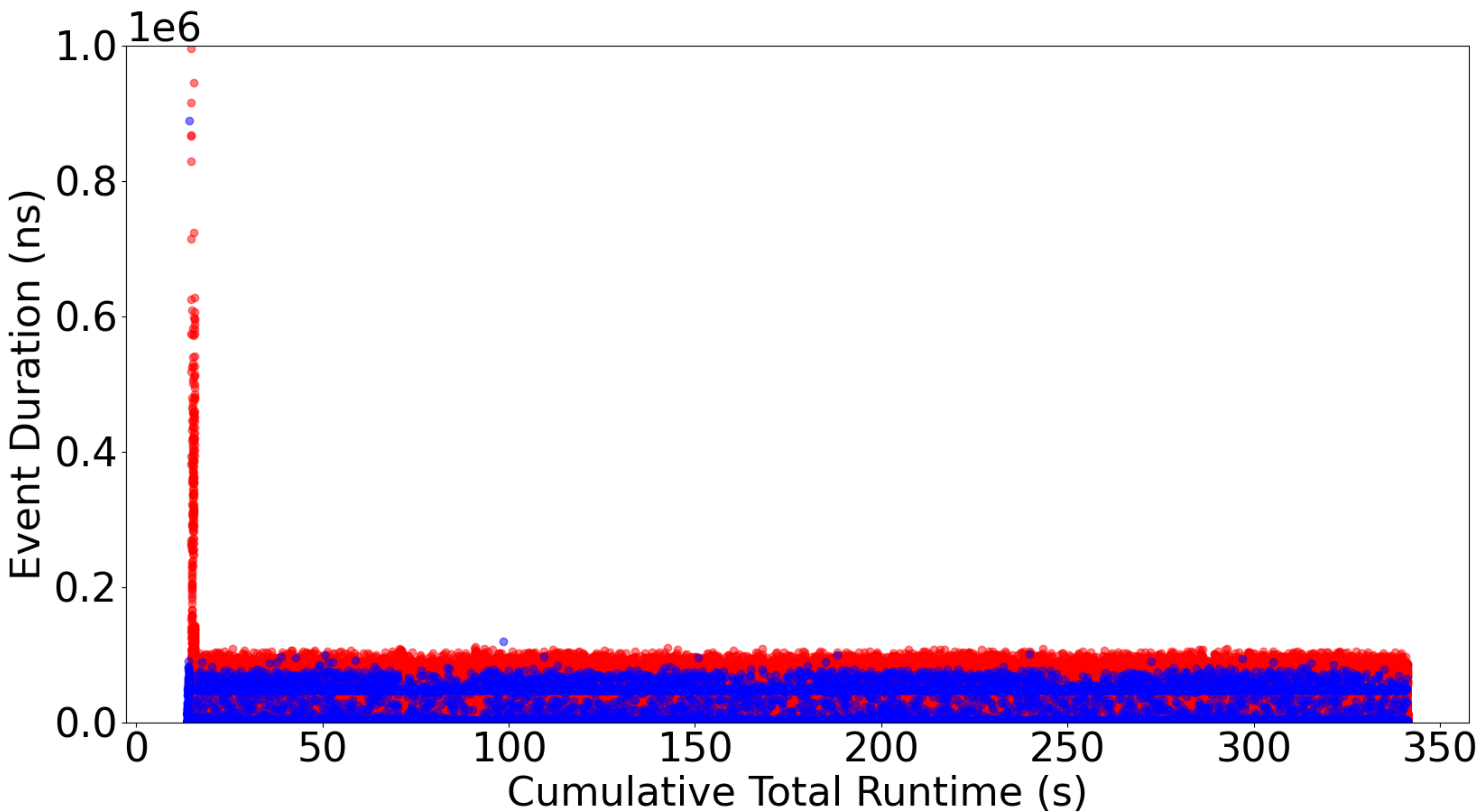}
    	\caption{DenseNet-201 Kernel and Transfer Times (Time-Slicing)}
    	\label{fig:transfer_slice_dense}
    \end{subfigure}
    \caption{Kernel execution times (red) and memory transfer operation times 
    (blue) for the DenseNet-201 inference task in both the baseline and 
    time-slicing scenarios.}
\label{fig:transfer-dense}
\end{figure*}

Time-slicing performed much worse with the RNNT training task for both BERT and
ResNet-34 than it did with the Pytorch training and inference combinations, as 
seen in Figure~\ref{fig:rnn_tt_util}. One reason for this is that time-slicing 
tends to perform worse when the tasks take longer due to having to perform more
context switches overall, and all three of these tasks were longer-running than
the PyTorch models were on average. However, in addition to this, ResNet-34 
possessed some attributes that would increase execution times by greater 
amounts when run concurrently with another application. The ResNet-34 inference
task spent orders of magnitude more time on memory transfers than other models 
performing inference did. Figure~\ref{fig:transfer-res} shows the kernel 
execution times and memory operation times of the Resnet-34 inference task in 
both the baseline and time-slicing cases, and Figure~\ref{fig:transfer-dense} 
shows the same for the DenseNet-201 inference task. ResNet-34 showed a 
significant increase in the amount of time spent on memory transfer tasks 
during the time-slicing case, while DenseNet-201 did not, suggesting that
memory transfer interference contributed to its higher turnaround times. These 
results align with previous findings that applications run as separate 
processes on NVIDIA devices can experience interference from memory transfer 
commands, despite being isolated as separate processes~\cite{hierarchy}. Like 
Observation~\ref{obs:contextswitch}, this is another way in which time-slicing 
does not actually isolate the two processes from each other.

\subsection{Multi-Process Service}

MPS allows applications run as separate processes to execute on a GPU at the 
same time. An MPS server is responsible for scheduling kernels from each 
process to the GPU. This differs from time-slicing in that the thread blocks of
kernels from separate processes can spatially share the GPU, i.e., execute on
the GPU at the same time, possibly even sharing an SM. While spatial sharing is
also possible when using priority streams, MPS allows the kernels to be from 
separate processes\footnote{To be more precise, they are launched from separate
CUDA contexts.} but does not include any notion of task prioritization. 
Instead, the MPS server can be configured to limit the number 
of threads that can be used by any one application;
for example, the 
MPS server can be set so that each client can use no more than 50\% of the 
total amount of threads offered by the GPU. NVIDIA recommends that this limit 
be set to 100\%/0.5\emph{n}, where \emph{n} is the number of clients to allow 
the GPU to potentially colocate kernels from separate applications on the same 
SM whenever there are idle resources. MPS is perhaps best suited to cases where
the kernels utilize less than the total available resources of the GPU. We 
implemented application-level concurrency by launching an MPS server and then 
launching both the inference and training tasks as separate MPS client 
processes.

In our experiments, we set the thread limit for both applications to 100\%. In
addition to this being the recommended setting, limiting the training 
application to only using some portion of the threads at any given time would 
defeat the purpose of using the training task to utilize spare GPU resources 
whenever they are available. 

\begin{obs}[O\ref{obs:contention}] \label{obs:contention}
While MPS increased utilization overall, it also caused intra-SM resource 
  contention that added to the execution times of both the training and 
  inference tasks.
\end{obs}

MPS saw consistent results in terms of utilization, as measured by the training
execution time in Figure~\ref{fig:util}. The additional 5000 
inference requests increased the time it took to train the model, usually by 
20-30 seconds. In contrast, using priority streams frequently increased the 
training task execution time by 30-40 seconds and using time-slicing by up to 
50 seconds. MPS can achieve good utilization primarily because it makes it 
possible to colocate blocks from different kernels. Priority 
streams was also able to colocate blocks, but it still took longer for the 
training task to complete because the inference task would be prioritized.

MPS additionally performs better when potential contention due to colocation is
low. Colocation of blocks from different applications allows for finer-grained 
resource assignment, but it can also create contention for resources when the 
blocks that are sharing an SM require the same resource, leading to significant 
performance degradation. Furthermore, unless it is clear what effects 
contention will have on the runtimes of the kernels, it is challenging to 
predict the performance of colocated kernels~\cite{perf, warped-slicer}. The 
increases in turnaround times compared to time-slicing observed in 
Figure~\ref{fig:turn} are partially explained by the presence of this resource 
contention. Minimizing contention is thus important for MPS to achieve 
increased utilization with less significant degradation in turnaround times.

\begin{obs}[O\ref{obs:balance}] \label{obs:balance}
MPS balances between the progress of the training and inference tasks, but it 
  is unable to adequately prioritize one over the other. More of the 
  degradation is seen on the part of the inference task due to the scheduling 
  policies used.
\end{obs}

While MPS can allow two kernels to spatially share the GPU, it is not able to
explicitly prioritize the execution of one application over another. Thus, both
the training and inference tasks are likely to make progress that is more 
balanced between the two applications than with priority streams. This is the 
main reason that the MPS training task execution times in 
Figure~\ref{fig:tt_util} are typically better than the priority streams times. 
Given that at least half of all of the models' inference and training kernels 
are small, MPS can employ this load-balancing during a significant portion of 
the tasks' execution, and there are frequently enough leftover resources from 
one application to share with the second.

The extent to which the GPU can be shared between applications is limited by 
the thread block scheduling policy. Kernels are still scheduled on essentially 
a first-come, first-served basis (up to the thread limit for each process). 
More specifically, our experiments suggest that the blocks are scheduled 
according to the \emph{leftover policy}, which dictates that all of the blocks 
from the most recently-arrived kernel must first be dispatched and executed on 
the GPU before any other kernels' blocks can be 
scheduled~\cite{scheduler-details}. Unlike priority streams, all of the blocks 
of the current kernel will be scheduled, and a later-arriving kernel is not 
able to schedule any before that. This presents a problem for the kernel that 
arrives at the GPU later, especially if that kernel is from the inference task,
as the running time of the second task is needlessly throttled. 

Due to these issues, MPS causes a greater degree of degradation for the 
inference tasks than the training tasks in Figure~\ref{fig:tt_util}. For 
instance, ResNet-152 saw the turnaround time increase 2X, but the training task
execution time only increased by a few seconds, which was the shortest training
task time observed for the three mechanisms. The DenseNet-201 training and 
inference tasks both had, on average, smaller proportions of long-running 
kernels with larger grid sizes. This was the model where MPS performed the best
in terms of both turnaround time, with an increase of 9.6ms, and utilization, 
which increased by only 11 seconds. For the other four Pytorch models that 
averaged longer-running training and inference kernels, the inference task was 
more often starved for resources, forced to make progress with what was 
leftover.

Unlike the Pytorch training tasks examined, RNNT had virtually no
large kernels during its runtime, meaning that there was almost always space on
the GPU for other tasks to use more of the resources. This is one reason why
MPS tended to perform more consistently well in terms of turnaround time than 
with the Pytorch models, which consisted of training tasks that more frequently
occupied the GPU with large kernels. However, RNNT's execution time increased 
more drastically than the Pytorch training tasks' execution times did using 
MPS, as seen in Figure~\ref{fig:rnn_tt_util}, due to the high amount of large 
kernels and the longer runtimes of the inference tasks.

Note that MPS's resource assignment issue is distinct from the compounded delay
problem discussed in Section~\ref{sec:prio_characterization}. The latter is 
caused by the gap between kernel launches where the training kernel uses the 
free resources. However, with a 100\% thread limit, the compounded delay 
problem is also an issue for MPS.
Thus, colocation and compounded delay also caused variance in turnaround time, 
as seen in Figure~\ref{fig:var-mps}. This variance was not as large as that 
observed in the priority streams case, as inference request satisfaction is 
partially dependent on the degree to which the training task is utilizing the 
GPU's resources.

 \section{Discussion}
\label{sec:observations}

\begin{figure}[t!]
    \centering
    \includegraphics[width=\columnwidth]{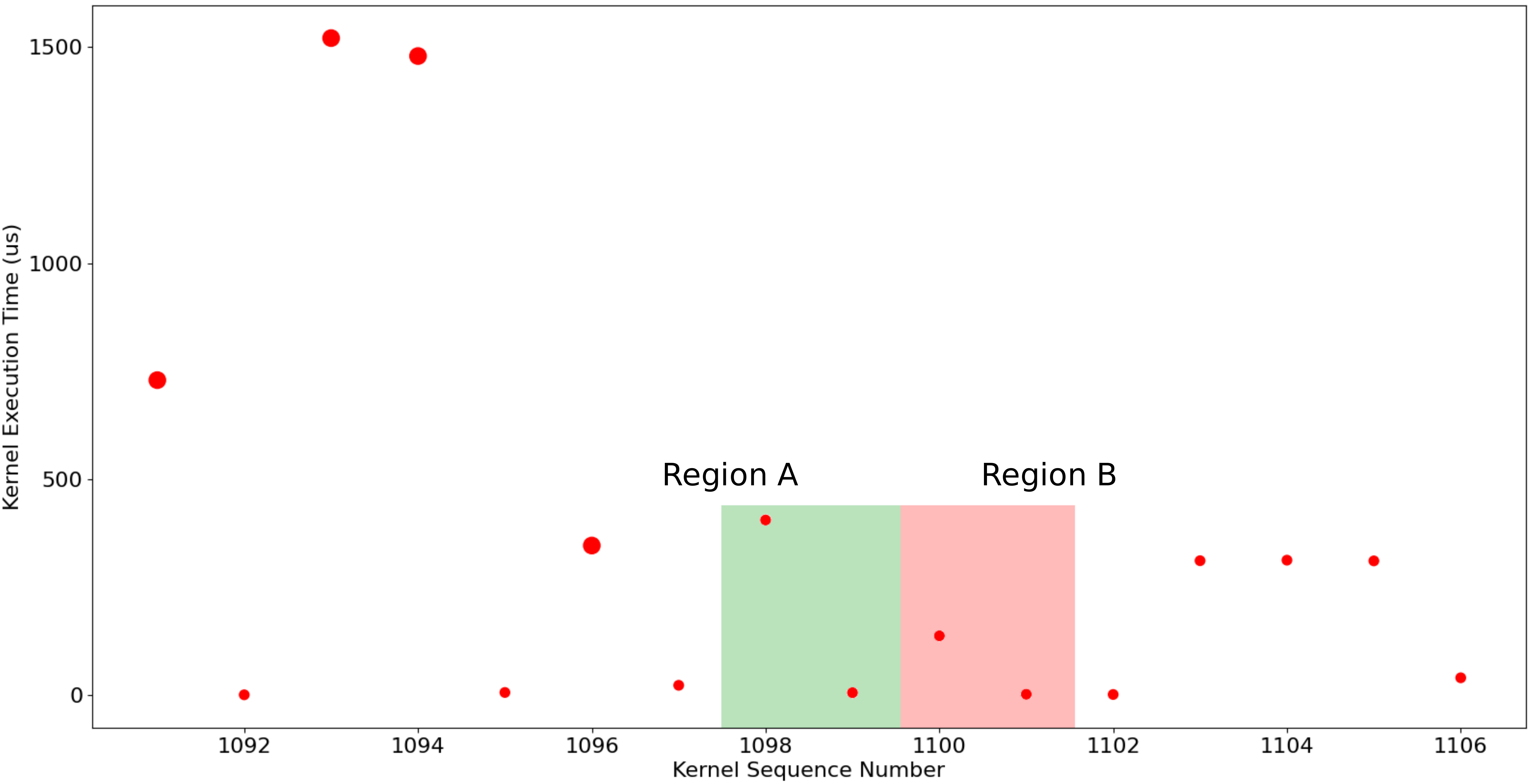}
    \caption{ResNet-152 Kernel Trace. A subset of the sequence of kernels 
    executed during the ResNet-152 inference task. The larger points are large
    kernels (in terms of their resource requirements on the GPU, they cannot
    fit their entire grid on the GPU at once), while the smaller points are small 
    kernels.}
    \label{fig:trace}
\end{figure}

\begin{obs}[O\ref{obs:preemption}] \label{obs:preemption}
For concurrent deep learning workloads, GPU utilization and predictability 
  could be improved with fine-grained preemption of thread blocks.
\end{obs}

Kernels which do not leave enough space for other kernels to be co-located 
alongside it often leave non-limiting resources unoccupied. Frequently, they 
prevent further blocks from being scheduled without actually utilizing all of 
the GPU's resources, when a different arrangement of thread blocks on the GPU 
would result in fewer of them being left idle.
 
Consider a case such as that of ResNet-152. For its training task, most of its 
large kernels were limited by threads. When such a training kernel is 
scheduled, the thread block scheduler places as many blocks as possible onto 
the GPU, but there will still be unused registers and memory resources.
The ResNet-152 inference task, in contrast, consists almost entirely of 
small, short-running kernels. Replacing even one block of the training kernel 
would leave room for a number of these smaller and less resource-intensive 
thread blocks. The combination of blocks from the training and inference 
kernels would use up an equivalent number of threads, but fit more blocks onto 
each SM and make better use of the registers and memory resources.

However, 
even when a given training kernel is 
small enough to leave space for the newly-arrived inference kernel to fit onto 
the GPU alongside it without waiting, resource contention will still incur 
delays that will push back further kernel launches in the sequence. 
There are two major issues to be solved here. First, the stochastic nature of 
the inference requests causes both unpredictability and inefficient 
utilization due to being unable to rearrange or interrupt thread blocks when 
they are already on the GPU. Second, resource contention also increases the
unpredictability even when blocks are colocated on the GPU to improve 
utilization. Solving these problems requires flexible scheduling mechanisms;
in particular, approaches which treat the GPU as a black box and make only
application-level scheduling decisions will not be sufficient. 

Future GPU architectures could potentially address the above utilization and 
predictability challenges with a new thread-block-level scheduling mechanism we
term \emph{fine-grained preemption}. Specifically, we define fine-grained 
preemption as the ability of the thread block scheduler to interrupt an 
arbitrary set of thread blocks at any point during the blocks' execution and 
relaunch those blocks at a later time. 

\begin{obs}[O\ref{obs:cost}] \label{obs:cost}
For the specific case of concurrent deep learning workloads, the cost of 
  fine-grained block-level preemption could be offset by the potential 
  benefits. 
\end{obs}

Current NVIDIA GPUs do not support fine-grained block-level preemption. As 
discussed in Section~\ref{sec:mechanisms}, while using priority streams does 
allow for one kernel to interrupt another kernel that is in the middle of being 
executed on the GPU, it does not
actually preempt any of the blocks currently on the GPU, instead waiting for
them to finish execution before scheduling any blocks of the new kernel. MPS 
similarly has no mechanism for interrupting the execution of a block, and 
instead schedules on a first-come, first-served basis; this lack
of block-level preemption is one cause of the performance degradation seen in
Section~\ref{sec:mechanisms}. The compounded delay incurred as a result caused
the priority stream turnaround times to be 
comparable to that of MPS, which has no notion of priorities at all. 
Time-slicing, while able to preempt blocks in the middle of their execution, 
can only do so in a course-grained way. It is
only able to clear the entire GPU of all currently-executing thread blocks, 
with no ability to partially preempt the GPU or prioritize one application over
another.

Fine-grained preemption would complement priority streams and MPS, and could 
address many of these issues which appear in the examined deep learning 
workloads. For instance, if enough blocks could be preempted as soon as an 
inference kernel arrived, none of the large or long-running training kernels 
would cause compounded delay. In addition, fine-grained preemption would allow 
MPS to prioritize the inference task over the training task. With the ability 
to clear a specific amount of space on the GPU at any time, MPS could include a
setting to specify a minimum resource usage requirement for each application, 
and preempt blocks to meet that threshold when the kernel arrives. 
Fine-grained preemption would also vastly improve 
predictability over the priority streams and MPS approaches if it were used in 
conjunction with a contention-aware scheduling policy, as the effects of 
compounded delay and the leftover policy could be eliminated.

The performance cost of fine-grained preemption depends on the implementation. For 
instance, a reasonable estimate for the cost of a full context-switch would 
consist of the time it would take to move the entirety of a kernel's context 
into global memory before the preempting kernel begins executing. This is 
because saving state is likely to be the dominating factor in preemption time. 
Using the methodology of prior work~\cite{chimera, deadline}, we estimate the 
cost for saving state when context-switching is 38$\mu$s. If all data from all 
82 SMs on the GPU need to be transferred to global memory, this would include 
64 KB of constant memory, 10496 KB of L1/shared memory, a 20992 KB register 
file, and 6144 KB of L2 cache data. With a total of 37696 KB to transfer to 
global memory, and a memory bandwidth of 936 GB/s~\cite{ampere}, saving state 
would take approximately 38$\mu$s to complete. Further, as global memory for
the target GPU is 24 GB, the storage size overhead is negligible.

It is not the case, however, that fine-grained preemption will necessarily
involve saving the state of the entire GPU. For a single SM, the context that 
needs to be saved includes 64 KB of constant memory, 128 KB of L1/shared 
memory, and a 256 KB register file, for a total of 448 KB. Assuming that an SM 
only has use of its fair portion of the bandwidth, it would have 11.4 GB/s of 
bandwidth to use. This results in a total of approximately 37$\mu$s. This is 
only 1$\mu$s less than the time it would take to save the state of all SMs.
Note that this estimate does not take into account 
factors such as maintenance of L2 cache coherence, memory transfer bandwidth 
interference from other applications, or the performance degradation likely to 
occur after switching due to reduced cache effectiveness.

Another method for estimating the cost of fine-grained preemption would be to 
examine the existing time-slicing mechanism. We conducted a simple experiment 
to estimate the amount of time between the last thread executed in time slice 
$n$ and the first thread executed in time slice $n + 1$. We launched two 
kernels as separate applications, with one thread block per SM each. 
Consequently, these kernels executed in alternating time slices. One thread in 
each thread block wrote the contents of the global timer register to local 
memory repeatedly, and we compared these time stamps across the two kernels to 
ascertain the amount of time between time slices. We observed an average time 
of approximately 145$\mu$s between recorded values; assuming half that time is 
spent saving the context of one kernel and the other half is spent resuming the
context of the other, the time to save state is 73$\mu$s.

Finally, we note that the existing time-slicing mechanism might serve well as 
the foundation for the proposed fine-grained preemption implementation. 
For example, it might be possible to reuse some of the same hardware for 
fine-grained preemption, reducing hardware costs. Further, time-slicing 
may already include techniques, such as only saving partial state, to reduce 
the performance cost of preemption. As we hypothesized in 
Observation~\ref{obs:contextswitch}, the NVIDIA RTX 3090 GPU does not appear to
transfer shared memory and registers when switching between contexts.

\begin{obs}[O\ref{obs:overlap}] \label{obs:overlap}
The cost of fine-grained block-level preemption can be hidden by taking 
  advantage of the fact that the deep learning tasks are a \textit{sequence} of 
  kernels. Preemption can be overlapped with transfer delay and the execution of 
  prior kernels in the sequence.    
\end{obs}

The trade-off for this lower and more predictable turnaround time is that the 
best-effort training task will take longer, by adding the overhead of 
preemption. However, the sequential nature of the kernels provides frequent
opportunities to hide the cost of preemption amidst transfer delay and the 
execution of other latter kernels. For example, data transfers from host to 
device take place periodically over the course of the applications' runtimes. 
Preemption for the kernels following such transfers can be hidden by 
performing some or all of the preemption during this transfer latency.

Preemption latency can also often be hidden for later kernels in the sequence
when a larger kernel follows a smaller one. While a smaller kernel is being 
executed on the GPU, since it is known that a larger kernel which requires more
resources will be following it, preemption cost can be hidden by preempting 
some of the blocks of the training task during 
the execution of the smaller kernel. This will guarantee that there will be enough space available to 
schedule the large kernel as soon as it arrives.
One such example is illustrated in Figure~\ref{fig:trace}, in the region labeled 
Region B. The first kernel only consists of 32 blocks of 64 threads each, while 
the second kernel has 512 blocks of 64 threads each. As the first kernel is 
being executed on the GPU, it will only take up 64 threads on 32 of the 82 SMs,
and the training task will use the rest of the resources. In order 
to make sure there is already room for the 512-block kernel when it arrives, the
training task can have enough blocks preempted during the execution of the first
kernel. The second kernel only takes 2$\mu$s to execute; on its own, preemption could
vastly increase the execution time of the kernel.
However, the first kernel takes about 137$\mu$s to execute, so it is able to hide
most of the delay that would be incurred if preemption occurs during its 
execution instead of after.

Preemption latency can also be avoided for the latter kernel in the same sequence by
simply leaving the space on the GPU open as the first kernel finishes, instead 
of filling it back up with training kernel blocks. An example where this could
be done can be seen in the 
region labeled Region A in Figure~\ref{fig:trace}. Both of the highlighted 
kernels are small, meaning that they can fit their entire grid on the GPU at 
one time; the first kernel has only 136 blocks of 256 threads each, while the 
second one has only 112 blocks of 32 threads each. However, the first kernel 
takes about 400$\mu$s to run, while the proceeding kernel
only takes around 6$\mu$s. This second kernel would be subsumed by the amount of 
time preemption would take; 
If the block placement of the first kernel is not ideal for the second
kernel, so replacing the blocks would result in an undesirable block placement
for the second kernel, the first kernel is also still long enough to hide the
latency of preempting a subset of training task blocks to create the ideal 
placement for the second kernel.

\begin{obs}[O\ref{obs:utilization}] \label{obs:utilization}
Utilization is difficult to define, but the execution time of the best-effort
  task is a good proxy metric.
\end{obs}

NVIDIA has a number of profiling tools available that offer different measures
of utilization. For instance, NSight Systems can report metrics such as SM 
ooccupancy and thread occupancy~\cite{nsys}, and \emph{nvidia-smi} and the NVML
API report
utilization as the percentage of the last one second where a kernel was running
on the GPU~\cite{nvml}.
Attempts to measure 
utilization with a single number often run the risk of over-simplifying the
actual state of resource usage due to both the number and variety of resources 
available. For example, if the current set of scheduled blocks uses all of the
available threads, this does not necessarily imply total saturation, as the 
amount of shared memory and registers in use could still be quite low. 
Measuring utilization as the execution time of the training task avoids this 
issue by taking into account how much useful work is being done by the 
background task with the resources it is occupying.

Take, for example, a single kernel from the ResNet-152 training 
task which has 100\% utilization based on thread usage as a metric: its grid consists of 200704
blocks, while each block consists of 256 threads. Thus, for our evaluation GPU, 
only 6 blocks can fit on each SM at a time, for a total of 492 blocks. The 
blocks of this particular 
kernel also use 32 registers per thread, for a total of only 49152 registers, 
meaning that the registers are being underutilized even when the GPU is fully 
occupied (as in, no more blocks can be scheduled because no more threads are 
available).

Now consider another example that also has 100\% utilization in terms of thread 
usage, but more completely saturates other GPU resources.
The average ResNet-152 inference kernel uses 2-3 warps per block; a typical 
example which comes up quite frequently during execution would be a 
convolutional implicit SGEMM kernel with 64 threads per block and 80 registers 
used per thread. Removing even one 256-thread block of the training task from 
an SM would make space for four blocks of this inference kernel, and register 
usage would now be at 61440. This is much closer to the limit of 64KB per SM, 
and the total number of blocks per SM is increased by 3, while the thread usage
still remains the same. Thus, we can see that a GPU having no more room to 
place another block does not mean that its resources are being utilized 
effectively, and that by merely rearranging the blocks and co-locating blocks 
from separate kernels, total resource usage can often be increased.

Both of these examples have 100\% utilization by the simple thread-based
utilization metric, but one utilizes more of the GPU's resources than the 
other. We do not necessarily solve this issue, but we find that the execution
time of the best-effort task is a good proxy measure to account for this 
situation.

 \section{Related Work}
\label{sec:related}

We divide our discussion of prior work into categories consisting of those 
studies which focus on reverse-engineering NVIDIA hardware, improving 
concurrency using spaital multiplexing, and improving concurrency using 
temporal multiplexing. Principally, we extend this prior work by evaluating
the performance of deep learning workloads under the existing methods for 
concurrent application execution NVIDIA GPUs.

\para{Reverse-Engineering Hardware.}
Previous work has characterized concurrency on NVIDIA hardware in terms 
of individual components of the NVIDIA scheduling hierarchy for general-purpose
workloads. For instance, previous work reverse-engineered the behavior of the 
hardware thread block scheduler~\cite{scheduler-details, streams} and 
investigated the performance implications of the observed thread block 
scheduling policies~\cite{perf}. For example, Xu et al. observed the thread 
block scheduler uses a leftover policy, where blocks from a later-arriving 
kernel are only scheduled once all earlier-arriving kernels' blocks have been 
scheduled first~\cite{warped-slicer}. The authors concluded this policy was 
often inefficient for concurrent application execution and performed very 
similarly to running two kernels in succession. We also observed issues with 
the leftover policy in Section~\ref{sec:mechanisms}.

Other work focuses on the higher levels of the scheduling hierarchy. Olmedo et 
al. discuss multiple levels of the NVIDIA scheduling 
hierarchy~\cite{hierarchy}, and the time-slicing application scheduler for the 
Tegra architecture similar to the one observed in this work is detailed by 
Capodieci et al~\cite{deadline}. The authors of the latter paper found the 
application-level scheduler to be lacking in the flexibility necessary to 
sufficiently prioritize latency-sensitive real-time tasks, and although their 
work was specific to an integrated architecture, we identify similar issues on 
the Ampere microarchitecture in Section~\ref{sec:mechanisms}.

\para{Spatial Multiplexing.}
Spatial multiplexing improves GPU utilization by efficiently dividing GPU
resources between kernels. 
Adriaens et al. proposed assigning each concurrent kernel exclusive access to a
subset of the SMs~\cite{spatial-multitasking}. In contrast, Xu et al. propose 
techniques that allow concurrent kernels to share SMs and attempt to optimize 
the placement of thread blocks from different kernels~\cite{warped-slicer}. The 
authors compare the performance of this technique directly to both the 
assignment of SM subsets to kernels and the leftover scheduling policy used by 
actual NVIDIA hardware when using streams. Pai et al. alternatively propose 
\textit{elastic kernels} which allow for fine-grained control over the
resources required by a kernel, and show its relative performance gains 
compared to CUDA streams with the leftover policy~\cite{elastic-kernels}. The 
above techniques assume a task set that is fixed and that all tasks are ready 
to be scheduled immediately. However, our work considers a stochastic deep 
learning workload in which the arrival times of the inference requests are 
unpredictable. An interesting direction for future work would be investigating 
how fine-grained preemption could be used in conjunction with the above 
spatial-multiplexing policies. In particular, with fine-grained preemption, the
scheduler can make the spatial-multiplexing decisions dynamically as new tasks 
arrive.

\para{Temporal Multiplexing.}
Temporal multiplexing improves the turnaround time of GPU 
applications rather than improving the utilization of GPU resources. 
Tanasic et al. proposed two forms of preemption for concurrent application 
execution on GPUs: context-switching and 
\textit{SM-draining}~\cite{enabling-preemption}. The latter is similar to 
priority streams in that it allows any currently-executing thread blocks to 
finish before replacing them with a different application's blocks. Park et al.
propose a third technique, \textit{SM-flushing}, in which blocks are 
interrupted and do not save their state, i.e., when resuming execution these 
blocks must start over from the beginning~\cite{chimera}. In addition, the 
authors built a scheduling framework that switches between these three 
preemption techniques dynamically. Capodieci et al. propose changes to the
time-slicing mechanism on the embedded NVIDIA Tegra architecture that would 
enable real-time task prioritization~\cite{deadline}.

All of the above techniques require hardware modification. In contrast, others 
have proposed higher-level methods to enable temporal multiplexing, such as 
kernel modifications to make GPU applications preemptable~\cite{flep}. Others
have examined reordering kernels or reordering memory transfer and kernel 
launch commands of an application~\cite{cumas, nimble}. 

\para{Deep Learning Workloads.}
Most of the above works also only consider general purpose GPU workloads. 
However, Xiao et al. examined a workload consisting of deep learning training
jobs. They proposed a scheduler for GPU server clusters which dynamically 
scales the memory and computational resources assigned to these jobs as their 
demand for them fluctuates and then schedules high-priority jobs and 
best-effort jobs cooperatively in the cluster through 
over-provisioning~\cite{antman}. However, this approach does not consider the 
microarchitectural interactions of the NVIDIA scheduling hierarchy such as 
the thread block scheduler which, as we have demonstrated, impact the 
performance of concurrent workloads.

Preliminary work conducted by Jain et al. on deep learning inference-only 
workloads suggests that combining spatial and temporal multitasking may 
outperform both in isolation~\cite{jain2018dynamic}. We discussed this 
possibility further in Section~\ref{sec:observations}.

 \section{Conclusions}
\label{sec:conclusion}

In summary, we have characterized three existing mechanisms for executing concurrent 
workloads currently available on NVIDIA GPUs---priority streams, 
time-slicing, and MPS---and their performance for handling concurrent 
deep learning workloads consisting of a best-effort training task and sequence
of latency-sensitive inference tasks. We considered their ability to provide
predictable and low turnaroaund times, while still maintaining high utilization,
and found that they each possessed certain drawbacks that made this difficult. 
Priority streams and MPS are both vulnerable to unpredictabile performance 
penalties incurred by resource contention and higher turnaround times due to 
the effects of compounded delay, while time-slicing lacked the spatial-sharing
capabilities to improve utilization significantly and showed evidence that 
memory transfer contention can sometimes interfere with maintaining low 
turnaround times.

We then argued that it is insufficient to consider only application-level 
scheduling, and the kernel- and block-level scheduling techniques such as 
fine-grained preemption would be necessary to efficiently execute concurrent
deep learning workloads. We showed that the deep learning workloads being 
examined have characteristics which make such a preemption mechanism necessary,
including sequential kernel launches, fluctuating resource requirements, and
stochastic arrival times. We additionally
demonstrated how these features present frequent opportunities to hide the cost
of fine-grained preemption, and also how such a mechanism could be used to 
complement the existing ones to increase utilization, turnaround time, and predictability. 
While the proposed preemption mechanism shows promise, testing 
fine-grained preemption on actual hardware will require modification to
proprietary NVIDIA components and, as such, cooperation from the NVIDIA
corporation. One potential direction for future work is to build on these findings by analyzing the 
performance of fine-grained preemption using a GPU simulator such as Accel-Sim~\cite{accel-sim}. 

We intend for this work to catalyze the creation
of more robust and efficient techniques for concurrent deep learning workloads
in the future. It is likely that such mechanisms should involve both 
efficient preemption mechanisms and 
contention-aware block placement policies to achieve greater concurrent 
workload performance. We also expect this work to serve as a baseline for 
comparison for work on concurrency mechanisms on NVIDIA devices. 
Additionally, we hope to 
see the proposed fine-grained preemption mechanisms implemented in future 
NVIDIA devices.

\bibliographystyle{ACM-Reference-Format}
\bibliography{bib}

\end{document}